\documentclass[aps,prd,twocolumn,twoside,superscriptaddress,floatfix, nofootinbib]{revtex4-1}

\usepackage{graphicx}
\usepackage{ifthen,amsmath,amssymb, bm}
\usepackage{graphicx}
\usepackage[dvipsnames]{xcolor}
\usepackage{hyperref}
\usepackage{float}
\usepackage{aas_macros}
\usepackage{caption}
\usepackage{subcaption}
\usepackage[export]{adjustbox}
\usepackage{bookmark}

\newcommand{\fnl}{\fnlloc}
\def\mpcoh{\,h^{-1}{\rm Mpc}} 
\def\gpcoh{\,h^{-1}{\rm Gpc}} 

\def\msunoh{\,h^{-1}{\rm M}_\odot}

\newcommand{\fnlloc}{f_{\rm NL}^{\rm loc}}

\begin{document}

\title{Refining local-type primordial non-Gaussianity: Sharpened $b_\phi$ constraints through bias expansion}

\author{Boryana Hadzhiyska}
\email{boryanah@berkeley.edu}
\affiliation{Physics Division, Lawrence Berkeley National Laboratory, Berkeley, CA 94720, USA}
\affiliation{Berkeley Center for Cosmological Physics, Department of Physics, University of California, Berkeley, CA 94720, USA}

\author{Simone Ferraro}
\affiliation{Physics Division, Lawrence Berkeley National Laboratory, Berkeley, CA 94720, USA}
\affiliation{Berkeley Center for Cosmological Physics, Department of Physics, University of California, Berkeley, CA 94720, USA}

\begin{abstract}
Local-type primordial non-Gaussianity (PNG), predicted by many non-minimal models of inflation, creates a scale-dependent contribution to the power spectrum of large-scale structure (LSS) tracers. Its amplitude is characterized by the product $b_\phi f_{\rm NL}^{\rm loc}$, where $b_\phi$ is an astrophysical parameter dependent on the properties of the tracer. 
However, $b_\phi$ exhibits significant secondary dependence on halo concentration and other astrophysical properties, which may bias and weaken the constraints on $f_{\rm NL}^{\rm loc}$. In this work, we demonstrate that incorporating knowledge of the relation between Lagrangian bias parameters and $b_\phi$ can significantly enhance PNG constraints. 
We employ the Hybrid Effective Field Theory (HEFT) approach at the field-level and a linear regression model to seek a connection between the bias parameters and $b_{\phi}$ for halo and galaxy samples, constructed using the \textsc{AbacusSummit} simulation suite and mimicking the luminous red galaxies (LRGs) and quasi-stellar objects (QSOs) of the Dark Energy Spectroscopic Instrument (DESI) survey.
For the fixed-mass halo samples, our full bias model reduces the uncertainty by more than 70\%, with most of that improvement coming from $b_\nabla$, which we find to be an excellent proxy for concentration.
For the galaxy samples, our model reduces the uncertainty on $b_\phi$ by 80\% for all tracers.  
By adopting Lagrangian-bias informed priors on the parameter $b_\phi$, future analyses can thus constrain $f_{\rm NL}^{\rm loc}$ with less bias and smaller errors. 

\end{abstract}
\maketitle



\section{Introduction} 
\label{sec:intro}

Understanding the origins of cosmic structures requires the investigation of the nature of primordial density fluctuations. One of the most exciting prospects of this quest is finding non-Gaussian features imprinted during the earliest moments after the Big Bang, which would illuminate the physics of the high-energy regime and put constraints on the widely accepted paradigm of inflation \citep{Bartolo:2004if}. Unraveling inflation would offer a unique path to probing the seeds of cosmic structure formation and allow us to delve deeper into the fundamental mechanisms of the Universe. 

Among the various types of primordial non-Gaussianity (PNG), the most well-studied is the local-type PNG, which links the small-scale with the large-scale galaxy density distribution and is characterized by a parameter $\fnlloc$. Local-type PNG holds significant promise for offering insights into the early universe, as a detection of $\fnlloc \sim 1$ would 
provide compelling evidence for the existence of multiple fields during inflation, which makes it a critical target for current and future cosmological surveys \cite{2014arXiv1412.4671A, 2022arXiv220308128A}.

A number of analyses have tried to constrain the parameter $\fnl$ both in the cosmic microwave background (CMB) \citep{Komatsu:2003iq} and in the large-scale structure (LSS) \citep{2008PhRvD..77l3514D}. While the Planck satellite has provided stringent constraints on  $\fnlloc$, with $\fnlloc = -0.9 \pm 5.1$ \cite{2020A&A...641A...9P}, future large-scale structure surveys are expected to significantly improve this sensitivity \cite{2017PhRvD..95l3513D,Sailer:2021yzm, 2018PhRvD..97l3540S,2023PhRvD.107f1301G, 2021JCAP...05..015M, 2022JCAP...04..013A, 2023EPJC...83..320J}, and bring it well beyond the forecasts for future CMB experiments \cite{2016arXiv161002743A,2019JCAP...02..056A}. The key to these improvements lies in the larger number of Fourier modes available for 3D vs. 2D surveys and the scale-dependent bias effect induced by local-type PNG on the galaxy density field on large scales. 

While theoretically exciting, there are still difficulties associated with observing the signal and interpreting it. For example, on large scales the effect of survey sample variance is extremely large. A particularly powerful approach for reducing this effect is multi-tracer analysis, which can potentially bring down the precision to $\sigma[\fnlloc] < 1$ \cite{2009PhRvL.102b1302S,2009JCAP...10..007M,2011MNRAS.416.3009B,2015PhRvD..91d3506F,2023EPJC...83..320J,2022JCAP...04..013A,2021PhRvD.104l3520D,2022JCAP...04..021M,2020MNRAS.498.3470W,2020RAA....20..158W}. Additionally, to uncover these subtle non-Gaussian imprints, we need to disentangle astrophysical and non-linear evolution effects from the features of the primordial Universe. These non-cosmological effects are quantified by the bias parameter $b_\phi$, which
a number of previous analyses assume to be perfectly correlated with the linear tracer bias $b_1$ (or halo mass) \cite{2008JCAP...08..031S, 2014PhRvL.113v1301L, 2021arXiv210613725M, 2019JCAP...09..010C, 2022PhRvD.106d3506C,2022arXiv220111518D}.

Recent studies have highlighted that the assumed relationship between $b_\phi$ and the linear bias $b_1$ might be oversimplified, especially when secondary halo properties related to the merger history are considered \cite{2020JCAP...12..013B, 2023JCAP...01..023L}. For this reason, some more recent works have instead reported constraints on the product $b_\phi \fnlloc$  \cite{2022JCAP...11..013B,2022arXiv220111518D,2022PhRvD.106d3506C}, which appears in the scale-dependent bias relation. This realization opens new avenues for constraining $\fnlloc$ by refining the models that relate $b_\phi$ to observable properties of galaxies and halos because any uncertainty on $b_\phi$ automatically translates into uncertainty on $\fnlloc$ as  $\sigma[\fnlloc] \propto \sigma[b_\phi]/b_\phi$. A promising approach involves the use of advanced simulations and bias expansion models to capture the complex dependencies of $b_\phi$.

State-of-the-art numerical simulations have revolutionized our ability to explore the complex interplay between primordial physics and LSS \citep[e.g.,][]{2009MNRAS.396...85D,2010JCAP...10..022W,2023ApJ...943...64C,2023ApJ...943..178C,2024JCAP...02..048F,2024A&A...689A..69G}. These simulations provide an invaluable laboratory for studying the link between galaxy formation and primordial physics, which is crucial for disentangling $\fnlloc$ from $b_\phi$. Recent works have explored the relationship between $b_\phi$ and $b_1$ in simulations \cite{2020JCAP...12..013B,2020JCAP...12..031B,2022JCAP...01..033B,2022JCAP...11..013B} and found that beyond the simple linear (universality) relation, there is a significant secondary dependence of $b_\phi$ on halo formation history properties such as concentration \cite{2023JCAP...01..023L,2024PhRvD.109j3530H}. Unfortunately, most of these properties are not observable, and instead, a number of large-scale analyses only measure `bias parameters' associated with a perturbative expansion model \citep{2004astro.ph.12025T}. Alternatively, machine-learning approaches trained on hydrodynamical simulations have shown to be able to capture some of the dependence of $b_\phi$ on secondary parameters and thus improve the constraints on $\fnlloc$, while at the same time allowing for a more optimal sample splits for multi-tracer analyses \cite{2023JCAP...08..004S}.

In this work, we demonstrate that incorporating a detailed understanding of tracer bias parameters into the estimation of $b_\phi$ can significantly enhance the precision of PNG constraints. A similar point was discussed recently in \cite{2024arXiv241018039S}, where the authors perform field-level fits of $b_\phi$ to perturbative mocks, highlighting the need for a reliable $b_\phi$ prior, due to the large residual degeneracy between $b_\phi$ and $\fnlloc$ even at the field level.

Here, we aim to refine the estimation of $b_\phi$ using a Lagrangian bias expansion model to fully non-linear N-body simulations and realistic galaxy samples. This approach allows us to incorporate the influence of various bias parameters and halo properties on $b_\phi$, thereby providing a more accurate and comprehensive understanding of the scale-dependent bias induced by local-type PNG. 
We utilize a suite of high-resolution N-body simulations, AbacusSummit, together with the `Separate Universe' technique, which can be employed to understand the response of small-scale galaxy physics to long-wavelength fluctuations.
Furthermore, we extend our study beyond halos to galaxies by utilizing an advanced halo occupation distribution (HOD) model to approximate the luminous red galaxies (LRGs) population targeted by the Dark Energy Spectroscopic Instrument (DESI) survey \cite{DESI:2019jxc}. We show that we are able to improve the uncertainty on $b_\phi$ and thus $\fnlloc$ by more than 80\% for all galaxy tracer samples considered in this study. 

We begin Section~\ref{sec:meth} by introducing the AbacusSummit simulation suite, describing relevant halo properties and the Separate Universe technique, detailing the construction of galaxy samples that mimic the galaxy survey DESI, and reviewing the theoretical background of the Lagrangian bias model HEFT. In Section~\ref{sec:res}, we analyze the dependence of $b_\phi$ on Lagrangian bias and concentration and present our forecasts for reducing the uncertainty on $b_\phi$ and thus $\fnlloc$. Importantly, we perform a realistic study of the scatter of $b_\phi$ for galaxy samples and investigate the effect of adopting a simple linear model. Finally, we discuss the implications of our findings for future LSS surveys and summarize our conclusions in Section~\ref{sec:disc}.

\section{Methods}
\label{sec:meth}

\subsection{AbacusSummit}

\textsc{AbacusSummit} is a suite of cosmological $N$-body simulations designed to meet and exceed the Cosmological Simulation Requirements of the DESI survey \citep{2021MNRAS.508.4017M}. The simulations were run with \textsc{Abacus} \citep{2019MNRAS.485.3370G,2021MNRAS.508..575G}, a high-accuracy, high-performance cosmological $N$-body simulation code, optimized for GPU architectures and for large-volume simulations, on the Summit supercomputer at the Oak Ridge Leadership Computing Facility.

We use the \texttt{base} resolution boxes of \textsc{AbacusSummit}, each of which contains 6912$^3$ particles in a $2\gpcoh$ box, each with a mass of $M_{\rm part} = 2.1 \times 10^9\msunoh$. While the \textsc{AbacusSummit} suite spans a wide range of cosmologies, here we focus on a single realization of the fiducial cosmology (\textit{Planck} 2018: $\Omega_b h^2 = 0.02237$, $\Omega_c h^2 = 0.12$, $h = 0.6736$, $10^9 A_s = 2.0830$, $n_s = 0.9649$, $w_0 = -1$, $w_a = 0$), \texttt{AbacusSummit\_c000\_ph000} as well as two additional runs with a 2\% higher and lower values of $\sigma_8$, \texttt{AbacusSummit\_c112\_ph000} and \texttt{AbacusSummit\_c113\_ph000}. In particular, utilize the halo and particle catalogs \citep{2022MNRAS.509..501H} as well as initial conditions outputs. For full details on all data products, see Ref.~\citep{2021MNRAS.508.4017M}.

\subsection{Concentration definition}

The recent merger activity of halos leaves an imprint on the halo concentration  \citep{1997ApJ...490..493N,2002ApJ...568...52W,2014MNRAS.441..378L,2016MNRAS.460.1214L}, and these imprints persist over several dynamical times ($\sim$Gyr) \citep{2020MNRAS.498.4450W}. Halo concentration impacts not only the clustering of halos, but also the halo occupation distribution of galaxies, thus affecting the clustering of galaxies in nontrivial ways
\citep{2001MNRAS.321..559B,2014MNRAS.441..378L,2015ApJ...799..108D,2014MNRAS.441.3359D,2018MNRAS.474.5143M}. Recent  evidence in data suggests that galaxy samples in modern spectroscopic surveys may exhibit a preference for selection biases dependent on the host halo concentrations of galaxies \citep{2021MNRAS.502.3582Y,2024MNRAS.530..947Y}. This motivates the study of the effect of concentration on the bias parameter associated with local-type PNG, $b_\phi$, the main subject of interest of this article.

Similarly to Ref.~\citep{2024PhRvD.109j3530H}, we adopt the following proxy for halo concentration:
\begin{equation}
\label{eq:conc}
c = r_{90}/r_{25},
\end{equation}
following the recommendation of Ref.~\citep{2022MNRAS.509..501H}, where $r_{90}$ and $r_{25}$ are defined as the radii, within which 90\% and 25\% of the halo particles are contained inside a sphere centered on the halo center. For more details on the halo finder and virial mass definition, we refer the reader to Ref.~\citep{1998ApJ...495...80B} and Ref.~\citep{2022MNRAS.509..501H}.

\subsection{Local environment}

Another secondary parameter (beyond halo mass and linear bias) that affects both the halo and galaxy clustering is the environment of the halo. This effect has been studied extensively in the literature \citep{2007MNRAS.378..641A,2017A&A...598A.103P,
2018MNRAS.476.5442P,2018MNRAS.473.2486S}. A halo residing in a dense region is expected to contain more galaxies on average than a halo in an underdense region because halos in overdense regions experience more mergers, whereas those in underdense regions accrete mass more smoothly. 

The response of galaxy formation in different halo environments in the presence of local-type PNG is of particular relevance to our study. To define halo environment, we adopt the same formalism as Ref.~\citep{2020MNRAS.493.5506H}. Specifically, for each halo, we find all neighboring halos beyond its virial radius but within $r_{\rm max} = 5 \mpcoh$ of the halo center. We sum the mass of all neighboring halos, denoted by $M_{\rm env}$,  and define the environment parameter,  $f_{\rm env}$:
\begin{equation}
  f_{\rm env} = M_{\rm env}/\bar{M}_{\rm env}
\end{equation}
where $\bar{M}_{\rm env}$ is the mean environment parameter within halo mass bin $M$.

\subsection{Separate Universe approach}
\label{sec:univ}

We employ the separate universe technique \cite{McDonald:2001fe, Sirko:2005uz, Wagner:2014aka, Lazeyras:2015lgp} to measure the local-type PNG-induced bias, $b_\phi$, by invoking the equivalence between the response of galaxy formation to long-wavelength perturbations and the response of galaxy formation to changes in the background cosmology. The separate universe argument states that as long as the physics of galaxy formation acts on much smaller scales compared with the size of the long-wavelength perturbations, these long-wavelength perturbations effectively are equivalent to a universe with slightly modified background.
Thus, the formation of galaxies at fixed cosmology in some region of space embedded in a long-wavelength fluctuation is equivalent to the formation of galaxies in a modified cosmology. 

To access the quantity $b_\phi$, the modified cosmology needs to have a different amplitude of the primordial scalar power spectrum, $A_s$, or equivalently, the amplitude of linear fluctuations smoothed on a scale of $8\ {\rm Mpc}/h$, i.e., $\sigma_8$. Mathematically, the PNG-induced bias, $b_\phi$, is defined as 
\begin{eqnarray}
\fnl b_\phi \equiv \frac{{\rm d} \ln n_h(z)}{ {\rm d}\phi},
\end{eqnarray}
which can be expressed as: 
\begin{eqnarray}
\label{eq:bphidef}
b_\phi = 4 \frac{{\rm d} \ln n_h(z)}{{\rm d} A_s} = 2 \frac{{\rm d} \ln n_h(z)}{{\rm d} \sigma_8}
\end{eqnarray}
by noting that \cite{2008PhRvD..77l3514D,2008JCAP...08..031S}
\begin{eqnarray}
\label{eq:tildeAs}
\tilde{A}_s = A_s\left[1  + \delta A_s\right], \ {\rm with} \ \ \ \ \ \delta A_s = 4\fnl\phi_L ,
\end{eqnarray}
where $\phi_L$ is to be treated as a constant locally and denotes the amplitude of the long-wavelength potential perturbation. Thus, we can evaluate $b_\phi$ in a separate universe with a different value of $A_s$ ($\sigma_8$), following the same procedure as \citep{2024PhRvD.109j3530H}, which we summarize below. In practice, we typically generate simulations with the same initial seed as the fiducial box, but with an input power spectrum file multiplied by $\left[1  + \delta \sigma_8^2 \right]$. In observations, making this measurement directly is extremely challenging, but recent works have suggested a possible path forward that directly extracts the PNG signal from the galaxy density field \citep{2023PhRvD.107f1301G}.

As in \citep{2024PhRvD.109j3530H}, we use the `Linear derivative' \textsc{AbacusSummit} boxes, \texttt{base\_c112\_ph000} and \texttt{base\_c113\_ph000}, which have the same initial seed as the fiducial simulation \texttt{base\_c000\_ph000}, but a different value of $\sigma_8$: namely, $\sigma_8^{\rm high} = 1.02 \times \sigma_8^{\rm fid}$ and $\sigma_8^{\rm low} = 0.98 \times \sigma_8^{\rm fid}$, respectively. We estimate $b_\phi$ in the separate universe approach as 
\begin{eqnarray}
\label{eq:bphimeasure}
b_\phi(z) = \frac{b_\phi^{\rm high}(z) + b_\phi^{\rm low}(z)}{2},
\end{eqnarray}
where
\begin{eqnarray}
b_\phi^{\rm high}(z) = \frac{2}{\delta \sigma_8^{\rm high}}\Big[\frac{n_h^{\rm high}(z)}{n_h^{\rm fid}(z)} - 1\Big], \\
b_\phi^{\rm low}(z) = \frac{2}{\delta \sigma_8^{\rm low}}\Big[\frac{n_h^{\rm low}(z)}{n_h^{\rm fid}(z)} - 1\Big].
\end{eqnarray}
where $n_h^{\rm fid}(z)$, $n_h^{\rm high}(z)$ and $n_h^{\rm low}(z)$ is the mean number density of halos (galaxies) in the fiducial, high- and low-$\sigma_8$ simulations at redshift $z$.

\subsection{Hybrid Effective Field Theory}
\label{sec:heft}

In this section, we summarize our approach for estimating the higher-order bias parameters, used to describe the connection between biased tracers such as galaxies and halos, and the underlying matter field in Lagrangian Perturbation Theory (LPT). For a review of perturbation theory, see  Ref.~\cite{2002PhR...367....1B,2018PhR...733....1D,2020moco.book.....D}. This standard version of LPT has been extended in the context of the Effective Theory of Large-Scale Structure \cite{Baumann:2010tm, Carrasco:2012cv, Senatore:2014eva, 2009JCAP...08..020M} in Lagrangian space \cite{2014JCAP...05..022P, 2015JCAP...09..014V, Chen:2020fxs, Chen:2020zjt}.

In the Lagrangian picture, we work with infinitesimal fluid elements labeled by their initial (Lagrangian) positions $\bm{q}$. The dynamics is described by the displacement $\bm{\Psi}(\bm{q},\eta)$, generated by the gravitational potential and such that the Eulerian (comoving) positions $\bm{x}$ of the fluid element at some conformal time $\eta$ is $\bm{x}(\bm{q},\eta) = \bm{q} + \bm{\Psi}(\bm{q},\eta)$ \cite{2015JCAP...09..014V,2014JCAP...05..022P}. 

More recently, a hybrid approach, Hybrid Effective Field Theory (HEFT), that leverages the strength of LPT as an analytical theory and the capability of $N$-body simulations to solve the full non-linear dynamics has been proposed and shown to perform very well in both simulations and observations \citep{2020MNRAS.492.5754M,2021MNRAS.505.1422K,2021JCAP...09..020H,2023MNRAS.524.2407Z,2023MNRAS.520.3725P}.

We adopt this framework here, and thus write the dependence of the galaxy field along its trajectory as an expansion to second order in the initial conditions adopting the Lagrangian framework \cite{2009JCAP...08..020M,2015JCAP...11..007S,2014JCAP...08..056A,2016JCAP...12..007V}:
\begin{align}
\label{eq:bias_expansion}
  F(\bm{q}) =& 1 + b_1 \delta_L(\bm{q}) + b_2 \big(\delta_L^2(\bm{q})-\langle\delta_L^2\rangle\big) \\ \nonumber &+ b_s \big(s_L^2(\bm{q})-\langle s_L^2\rangle \big) + b_\nabla \nabla^2 \delta_L(\bm{q}) , 
\end{align}
where $b_1$, $b_2$, $b_s$ and $b_\nabla$ are free bias parameters and $s_L^2 = s_{ij} s^{ij}$ is the shear/tidal field, with $s_{ij}\equiv ( \partial_i \partial_i/\partial^2 - \delta_{ij}/3 )\ \delta_L$. We have also included the lowest-order non-local term $\nabla^2\delta_L$. The functional can then be advected to the real-space (Eulerian) position $\bm{x}$ \cite{2008PhRvD..77f3530M}:
\begin{equation}
    1 + \delta_{\rm g}(\bm{x}) 
   = \int d^3\bm{q}\,F(\bm{q})\, \delta^D(\bm{x}-\bm{q}-\bm{\Psi}(\bm{q}))
\label{eqn:deltag}
\end{equation}

To estimate the bias parameters, $b_i$, of our model, we minimize the difference between the model and the observable, $\delta_{\rm g}$. This is similar to the procedure adopted in \citep{Schmittfull:2018yuk, 2022MNRAS.514.2198K, Baradaran:2024jlh}, which we follow closely. In brief, the stochastic residual field $\epsilon(\mathbf{k})$, after removing deterministic contributions is given by:
\begin{equation}
\epsilon(\mathbf{k}) = \delta_{\rm g}(\mathbf{k}) - \delta_m(\mathbf{k}) - \sum_i b_i(\mathbf{k}) \mathcal{O}_i(\mathbf{k}),
\end{equation}
Minimizing the power spectrum of $\epsilon$, the bias parameters are given by:
\begin{equation}
\hat{b}_i(\mathbf{k}) = \left\langle \mathcal{O}_i \mathcal{O}_j \right\rangle^{-1} (\mathbf{k}) \left\langle \mathcal{O}_j(-\mathbf{k}) \left[ \delta_{\rm g}(\mathbf{k}) - \delta_m(\mathbf{k}) \right] \right\rangle .
\end{equation}
When fitting up to a maximum wavenumber, $k_{\rm max}$, this reduces to:
\begin{equation}
\hat{b}_i = M_{ij}^{-1} A_j.
\end{equation}
where $A_j$ and $M_{ij}$ are defined as
\begin{eqnarray}
A_j = \left\langle [\mathcal{O}_j(\mathbf{x}) (\delta_{\rm g}(\mathbf{x}) - \delta_m(\mathbf{x}))]_{k_{\rm max}} \right\rangle ,
\\ \nonumber
= \int_{{|\mathbf{k}| < k_{\rm max}}} 
\frac{d^3k}{(2\pi)^3} \mathcal{O}_j(\mathbf{k}) [\delta_{\rm g} - \delta_m]^*(\mathbf{k}) ,
\end{eqnarray}
and 
\begin{eqnarray}
M_{ij} = \left\langle [\mathcal{O}_i(\mathbf{x}) \mathcal{O}_j(\mathbf{x})]_{k_{\rm max}} \right\rangle ,
\\ \nonumber
= \int_{k_{|\mathbf{k}| < k_{\rm max}}} \frac{d^3k}{(2\pi)^3} \mathcal{O}_i(\mathbf{k}) \mathcal{O}_j^*(\mathbf{k}). 
\end{eqnarray}

Recently, in a similar vein to our work, a number of studies have adopted field-level estimates of the bias parameters associated with perturbation theory, 
to study various cosmological and astrophysical phenomena \citep[see e.g.,][]{2024arXiv241201888I,Shiferaw:2024ehr,2024arXiv241201888I}. Of particular relevance to our work are 
Refs.~\cite{2023JCAP...08..004S,2024PhRvD.110f3538I,2024arXiv241018039S}, which apply similar field-level bias estimation techniques to the study of PNG.

\subsection{Halo Occupation Distribution}
\label{sec:hod}

In this work, we use the same mock catalogs that were constructed in Ref.~\citep{2024PhRvD.109j3530H} based on fits to the LRG and QSO clustering with DESI 1\% data \citep[see][and Table 1 therein for the parameter values we adopt when creating our synthetic LRG and QSO catalogs]{2024MNRAS.530..947Y}. For our fiducial galaxy catalogs, we adopt the best fit values of these fits at redshifts $z = 0.5$, $z = 0.8$, and $z = 1.4$ for the three samples shown here, using the \textsc{AbacusSummit} boxes to ensure consistency of the halo mass and HOD parameter definitions. We then consider assembly bias extensions around these fiducial values, including both environment and concentration dependence.

DESI targets LRGs at $z \lesssim 1$, with a fairly constant number density between $0.4 < z < 0.8$ of approximately $5 \times 10^{-4} [{\rm Mpc}/h]^{-3}$. QSOs (quasars), on the other hand, are the tracer choice for studying large-scale structures at high redshifts, with a roughly constant number density between $0.8 < z < 2.1$, at $2 \times 10^{-5} [{\rm Mpc}/h]^{-3}$.

The HOD model that we adopt for both LRGs and QSOs is given by the formalism of Ref.~\citep{2005ApJ...633..791Z}:
\begin{equation}
    \bar{n}_{\mathrm{cent}}(M) = \frac{f_\mathrm{ic}}{2}\mathrm{erfc} \left[\frac{\log_{10}(M_{\mathrm{cut}}/M)}{\sqrt{2}\sigma}\right], \label{eq:zheng_hod_cent}
\end{equation}
\begin{equation}
    \bar{n}_{\mathrm{sat}}(M) = \left[\frac{M-\kappa M_{\mathrm{cut}}}{M_1}\right]^{\alpha}
    \label{eq:zheng_hod_sat}
\end{equation}
where $M_{\mathrm{cut}}$ determines the minimum mass of a halo to host a central galaxy, $M_1$ sets the pivot scale of the power law of satellite occupation, $\sigma$ controls the steepness of the transition from 0 to 1 in the number of central galaxies, $\alpha$ is the power law index on the number of satellite galaxies, $\kappa M_\mathrm{cut}$ gives the minimum halo mass to host a satellite galaxy, $f_\mathrm{ic}$, which is a downsampling factor controlling the overall number density of the mock galaxies.

The velocity of the central galaxy is taken as the average velocity of the so-called ``L2'' subhalo \citep[see][]{2022MNRAS.509..501H}. For the satellite galaxies, the velocities are inherited from random halo particles. The analysis of Ref.~\citep{2023arXiv230606314Y} also includes velocity bias, which is necessary for modeling redshift-space clustering on small scales \citep[e.g.][]{2015MNRAS.446..578G,2022MNRAS.510.3301Y}. For a detailed description of the galaxy population models, see Ref.~\citep{2024PhRvD.109j3530H}.

\section{Results}
\label{sec:res}

\begin{figure*}
    \centering
    \includegraphics[width=.24\linewidth]{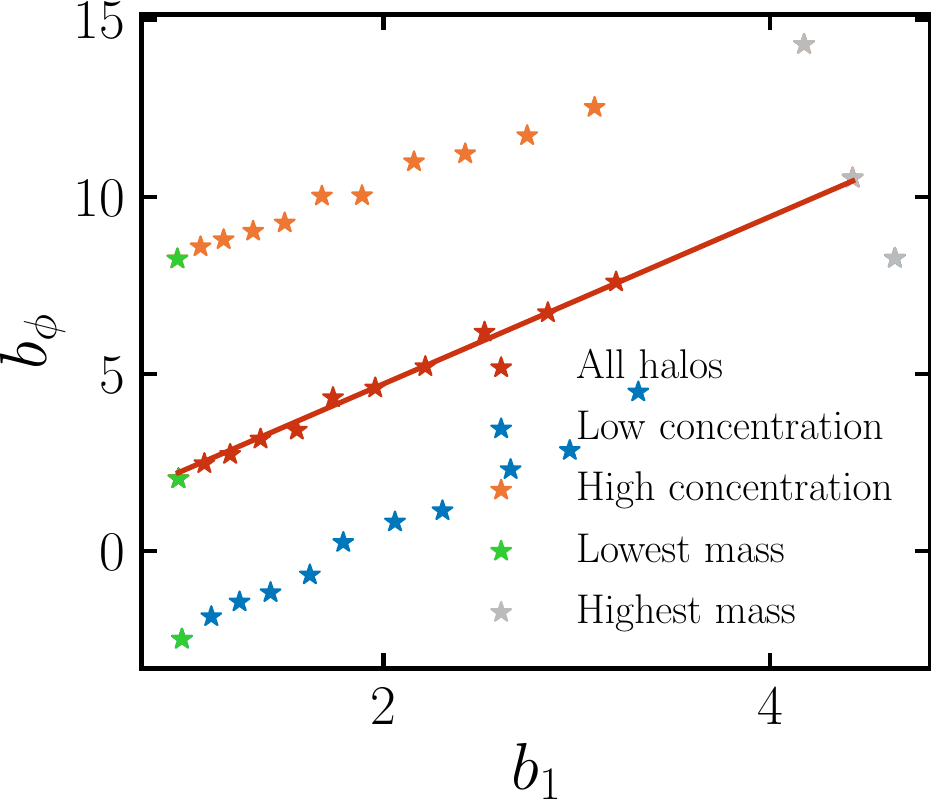}
    \includegraphics[width=.24\linewidth]{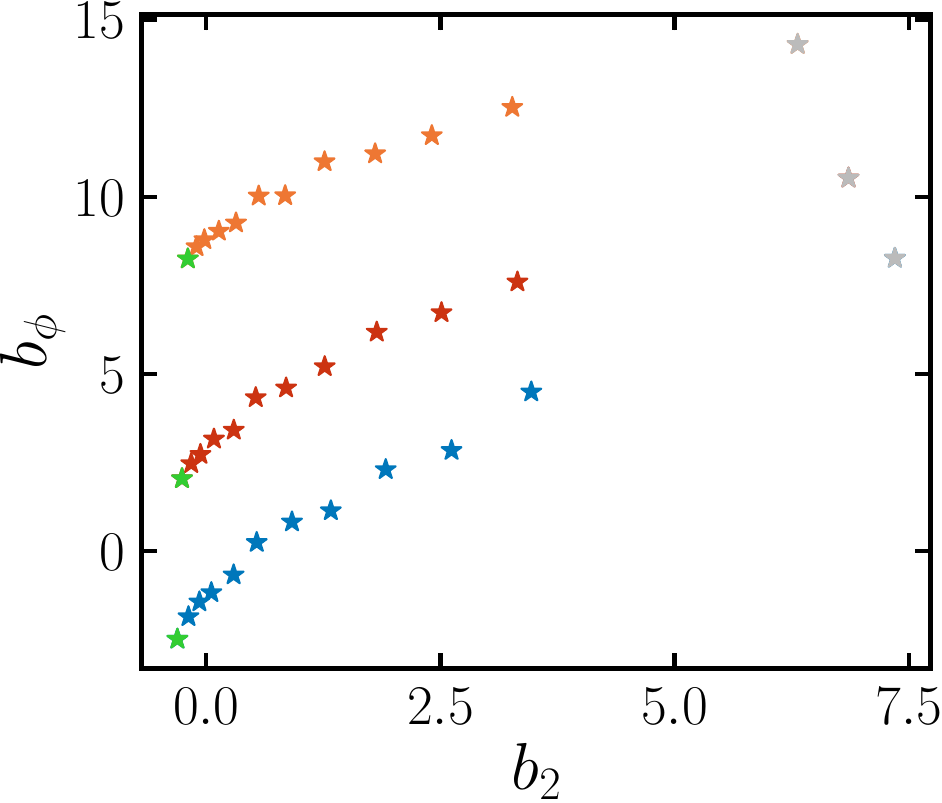}
    \includegraphics[width=.24\linewidth]{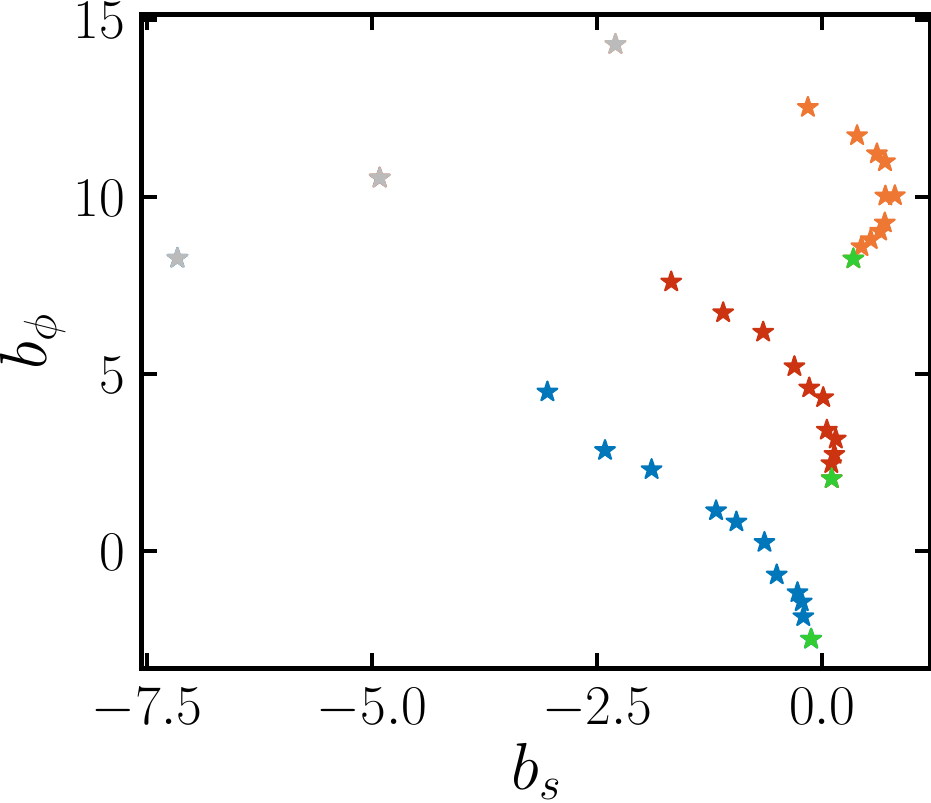}
    \includegraphics[width=.24\linewidth]{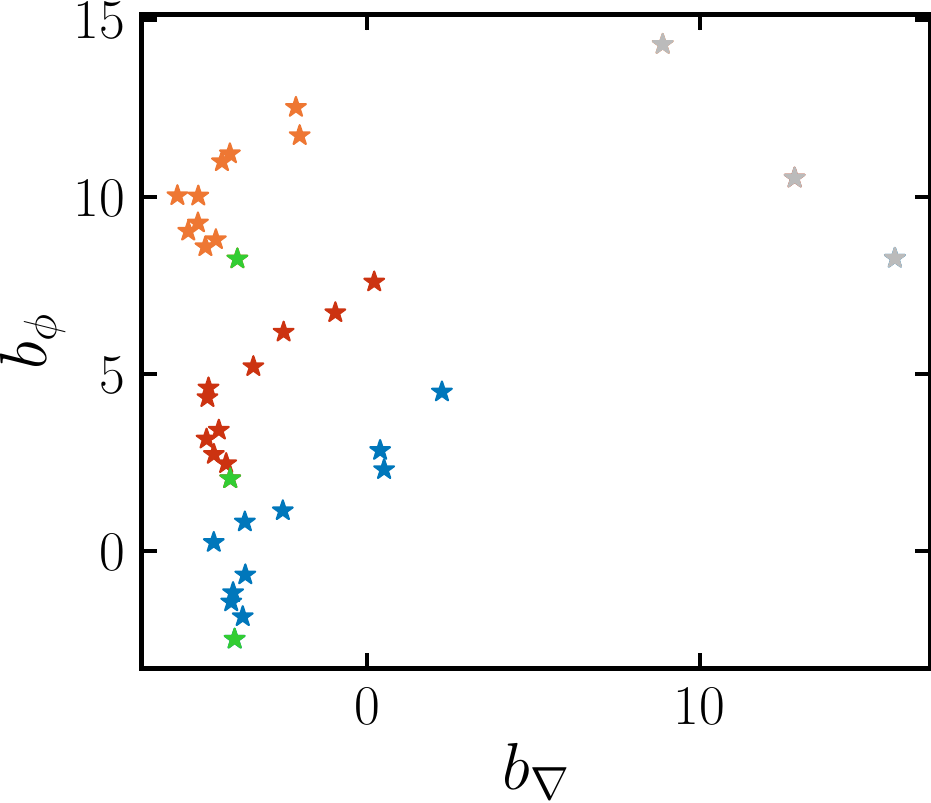}
    \caption{Universality relations for all bias parameters as a function of halo mass and concentration, for halos at $z = 0.5$ split into 17 mass bins between 12.5 and 14.5 and three equally sized concentration bins. The effect of concentration is mostly captured as a variation in the $p$ parameter (with $c = 0.7$, see Eq.~\ref{eq:univ}). Similarly to $b_1$, the relationship between $b_2$ and $b_\phi$ is almost exactly linear. $b_s$ exhibits an opposite trait, where large values of $b_s$ correspond to smaller values of $b_\phi$. Small values of $b_\nabla$ and large values of $b_s$, which are typically measured for low-mass halos, map to a large range of $b_\phi$, making that relationship less informative on its own.}
    \label{fig:halo_univ}
\end{figure*}

The focal point of this paper is the study of the relationship between $b_\phi$ and the Lagrangian biases, $b_X$, as well as the assembly bias parameters such as concentration and environment. We do this for both galaxy and halo samples, though we note that in realistic surveys halos are not directly accessible. Nonetheless, halo samples provide key insights and indicate the power of future cluster-based analysis.

\subsection{Halos}

\subsubsection{Universality relations and concentration}

When performing PNG analysis, many works assume that the $b_\phi$ parameter is known exactly, and its value is extracted from the well-known universality relation, linking it to the linear bias parameter, $b$, as \cite{2008PhRvD..77l3514D, 2008JCAP...08..031S}:
\begin{equation}
\label{eq:univ}
    b_\phi = 2 c \delta_c (b - p),
\end{equation}
where in this work, we adopt $p = 1$, $c$ ranges from 0.7 to 1, and $\delta_c = 1.686$. Note that the connection between linear bias and the Lagrangian bias, $b_1$, is given by $b = 1 + b_1$.

However, as exemplified by a number of recent works, the scatter between $b_\phi$ and $b_1$ is non-trivial and non-negligible, since $b_\phi$ exhibits strong dependence on various astrophysical properties of the sample. One of the well-known properties in the case of halos is concentration. Additionally, the other bias parameters, $b_2$, $b_s$, and $b_\nabla$, also encode astrophysical information and can thus help in pinpointing the value of $b_\phi$ better and reducing the scatter. First, we explore the universality relations for all bias parameters as a function of halo mass and concentration in Fig.

We split the halos at $z = 0.5$ into 17 mass bins between 12.5 and 14.5. At each mass bin, we split the halos into three equally sized concentration bins and compute the value of $b_\phi$ as well as the additional bias parameters, $b_1$, $b_2$, $b_s$ and $b_\nabla$, for each case. The relation between $b_1$ and $b_\phi$ follows the universality relation very closely, as expected. The effect of concentration is mostly captured as a variation in the $p$ parameter (see Eq.~\ref{eq:univ}), with the high-concentration halos exhibiting larger values of $b_\phi$ compared with the low-concentration ones. Similarly to $b_1$, the relationship between $b_2$ and $b_\phi$ is almost exactly linear, with concentration modulating that relation by a constant positive or negative offset to $b_\phi$, but with no significant shift to $b_2$. $b_s$ exhibits an opposite trait, where large values of $b_s$ correspond to smaller values of $b_\phi$. In addition, we see that the value of $b_s$ at small halo mass (i.e. large $b_s$) seems to be very stable. In other words, given a measurement of $b_s$ close to zero for some halo sample, one would not be able to pinpoint very well what the value of $b_\phi$ is for those halos. Similarly, small values at $b_\nabla$, which are typically measured for low-mass halos, map to a large range of $b_\phi$, making that relationship less informative, provided that only $b_\nabla \approx -5$ is measured for a given halo sample.

However, in practice we typically know the halo mass or equivalently linear bias fairly well for a given halo (or galaxy) sample. Thus, a more interesting question is whether measuring these additional bias parameters can help us reduce the scatter at fixed halo mass. We reiterate that while these universality relations give us an idea of the behavior of these halos on average at some fixed mass, they do not capture deviations due to selection effects such as concentration assembly bias. To better understand how well we can characterize the scatter around the mean values of $b_\phi$ at fixed halo mass, we study the correlation between $b_\phi$ and the 4 bias parameters at fixed halo mass.

\subsubsection{Correlation between bias parameters}

\begin{figure*}
    \centering
    \includegraphics[width=.195\linewidth]{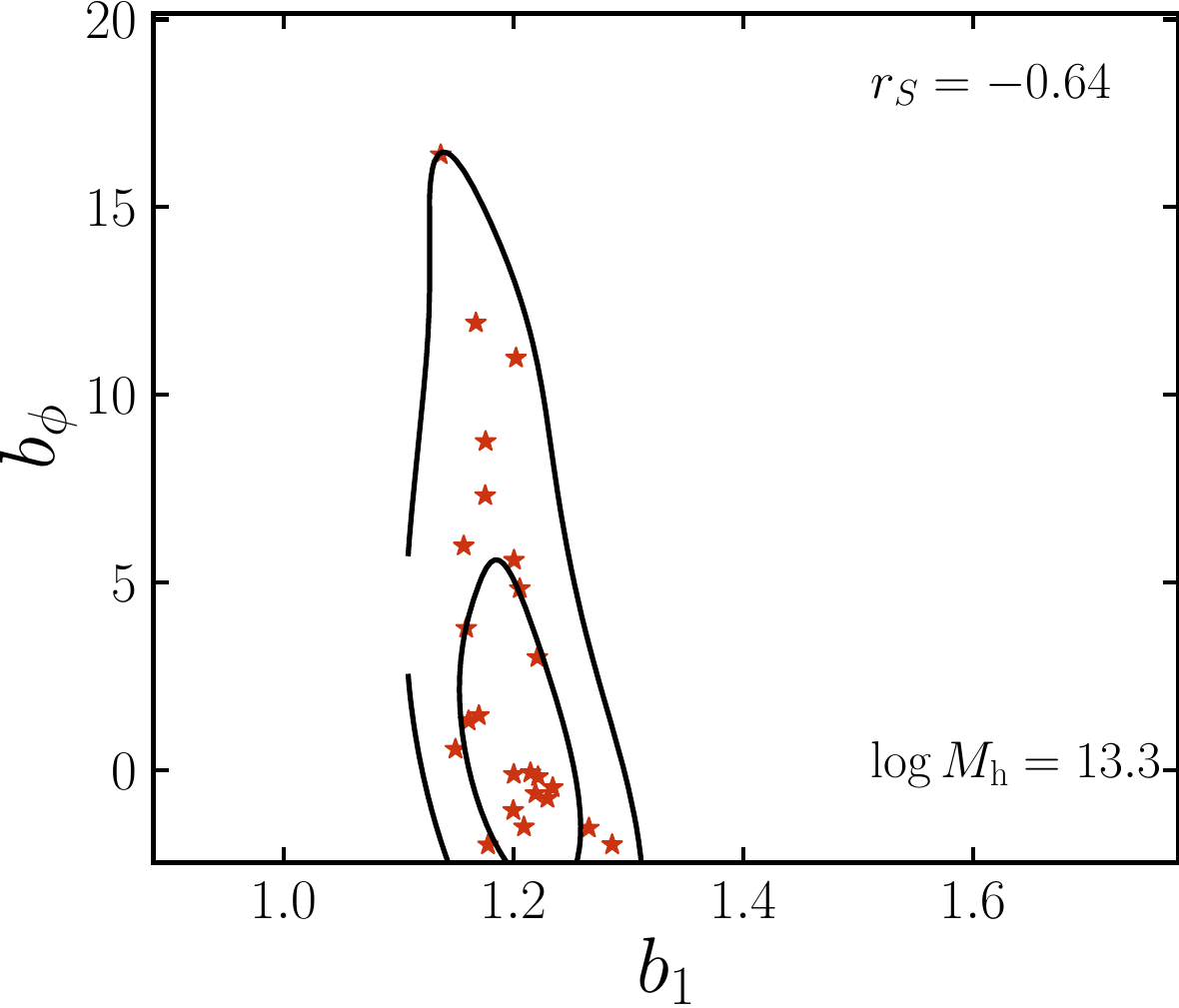}
    \includegraphics[width=.195\linewidth]{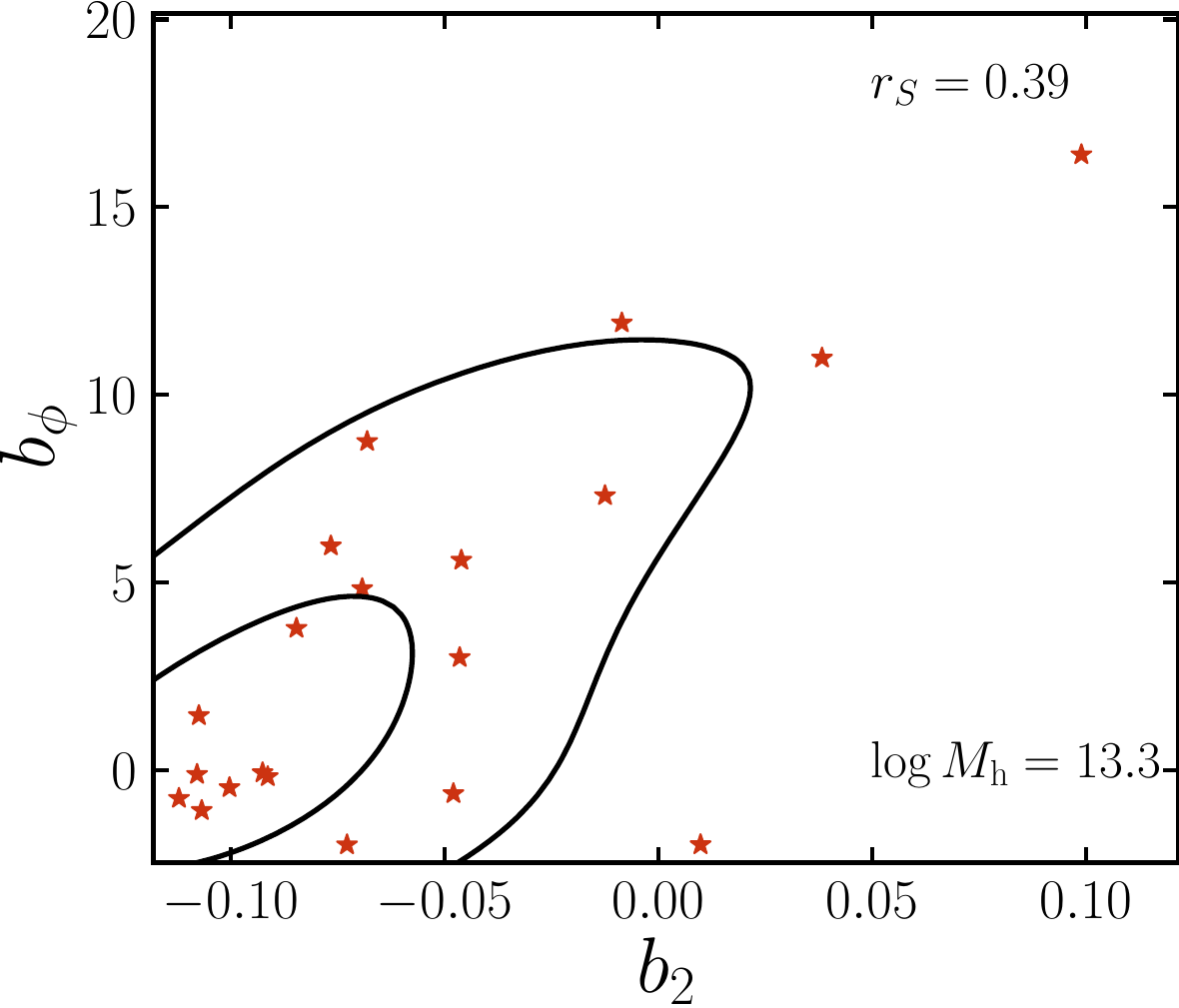}
    \includegraphics[width=.195\linewidth]{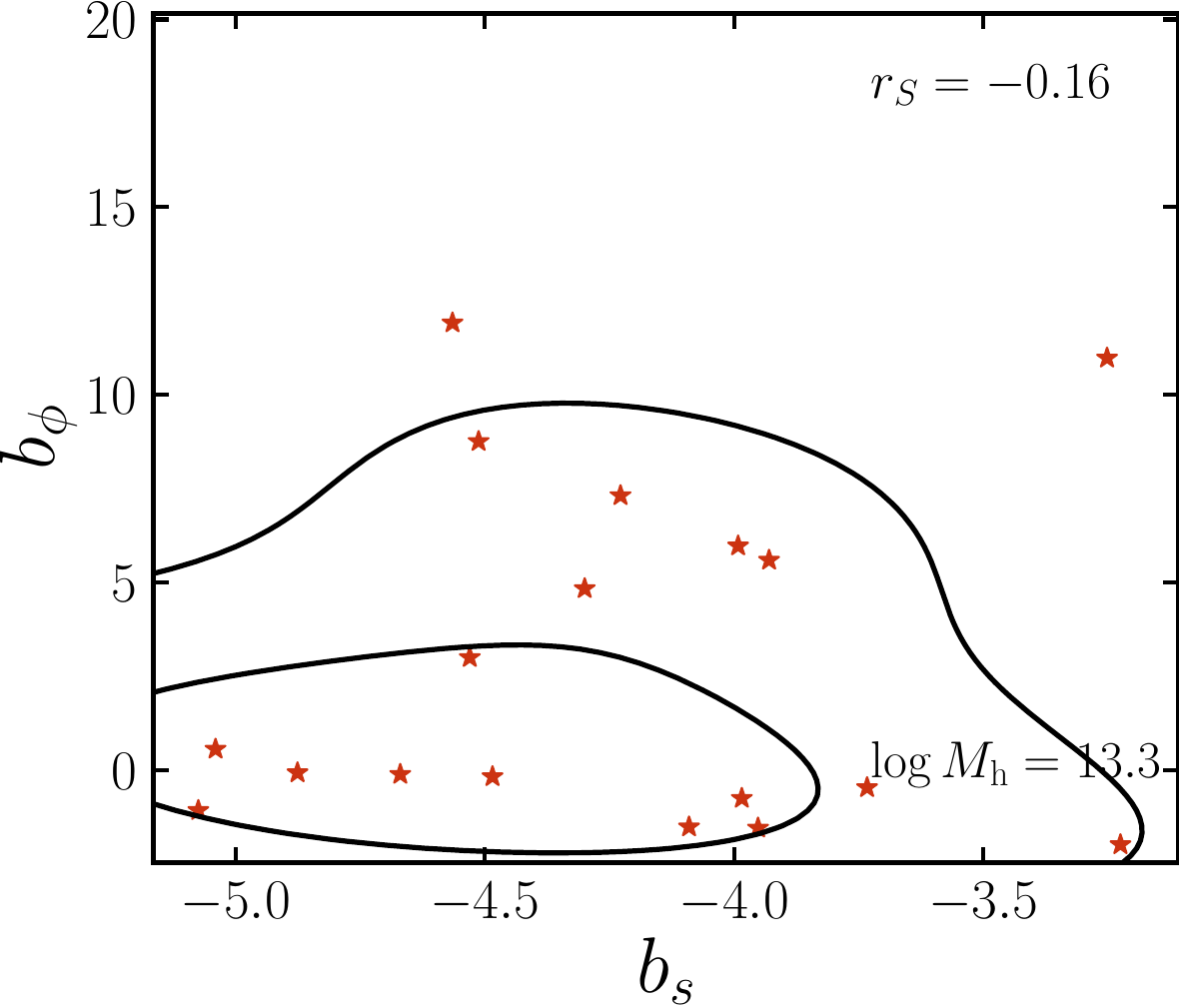}
    \includegraphics[width=.195\linewidth]{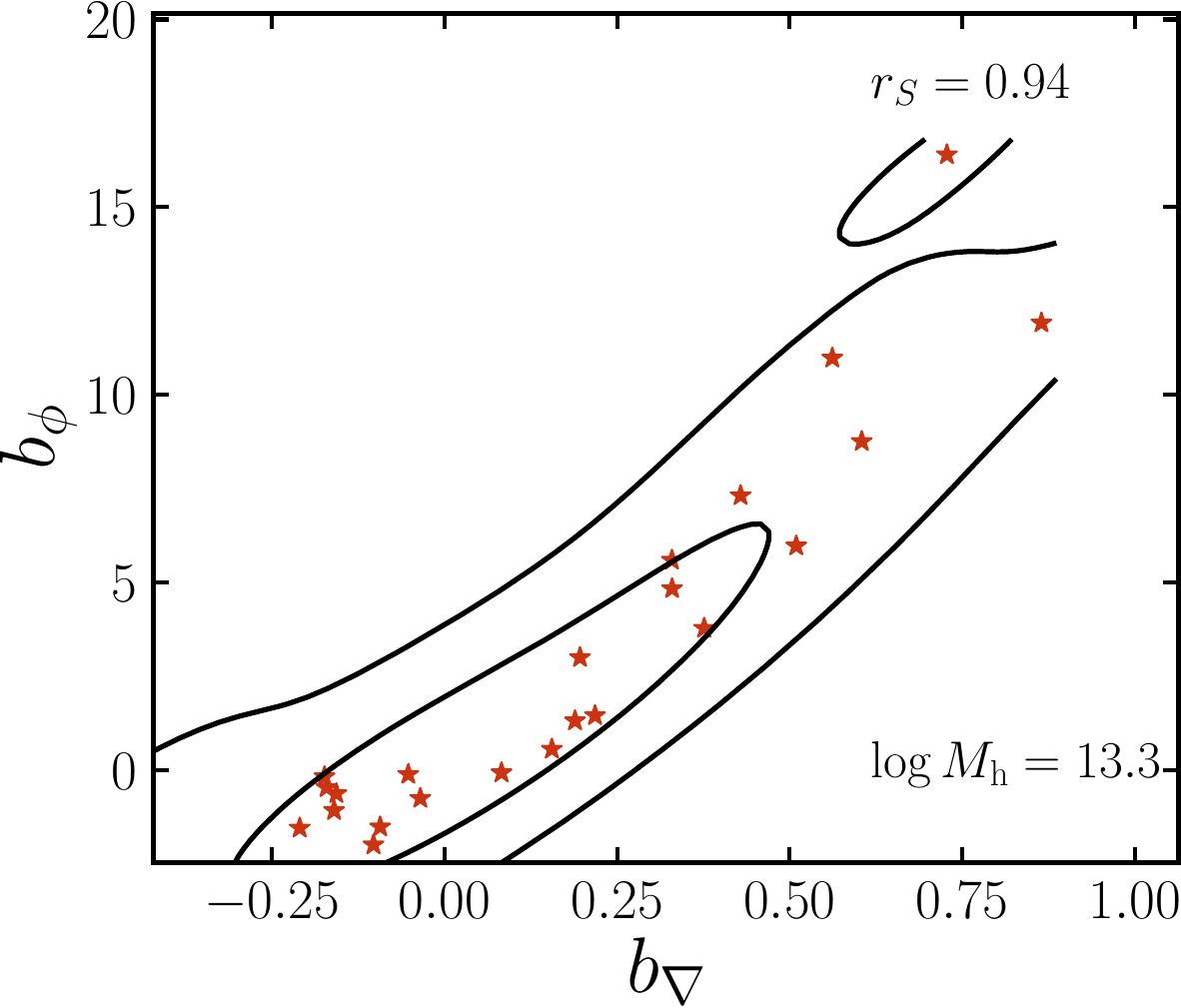}
    \includegraphics[width=.195\linewidth]{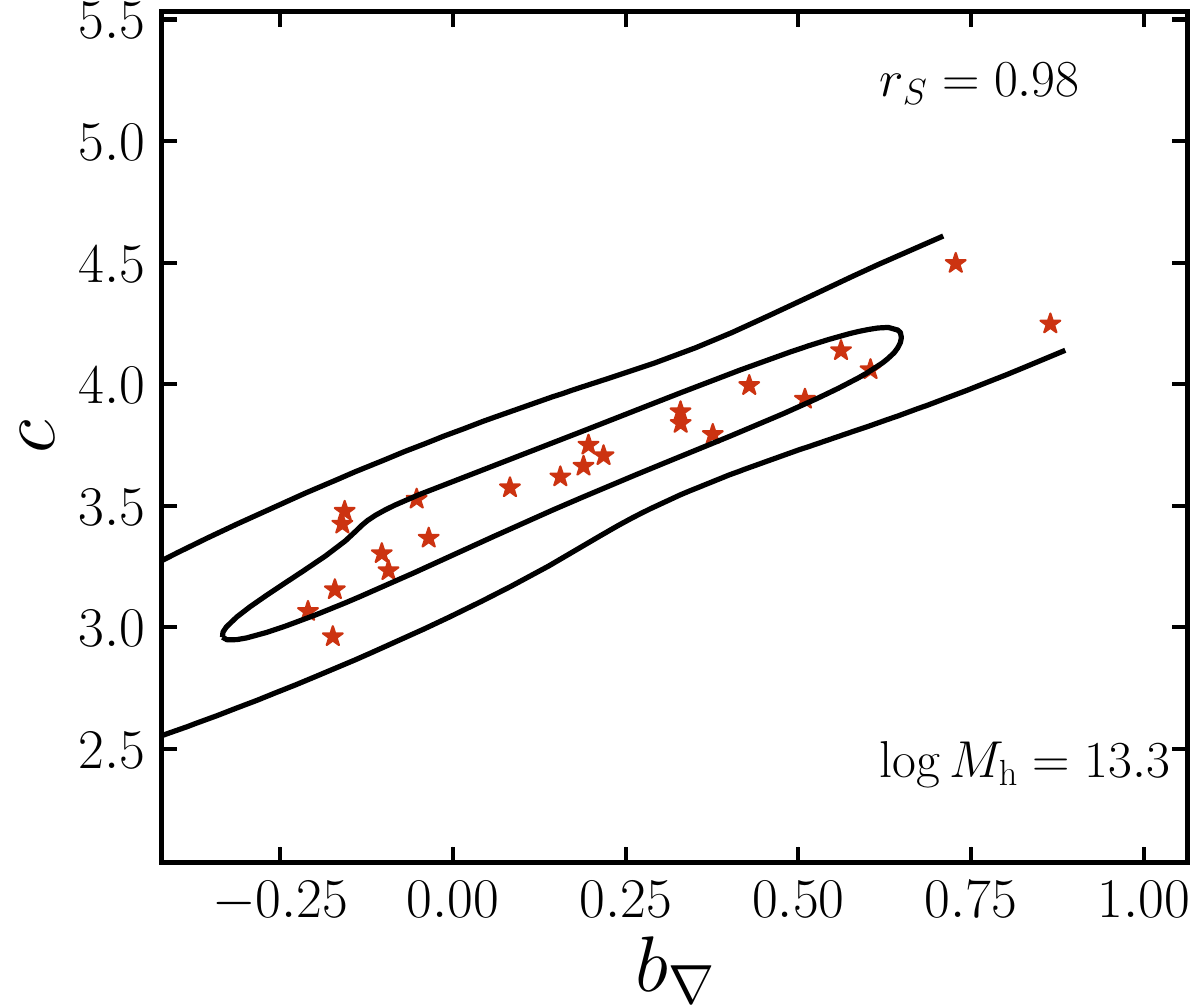}
    \\
    \includegraphics[width=.195\linewidth]{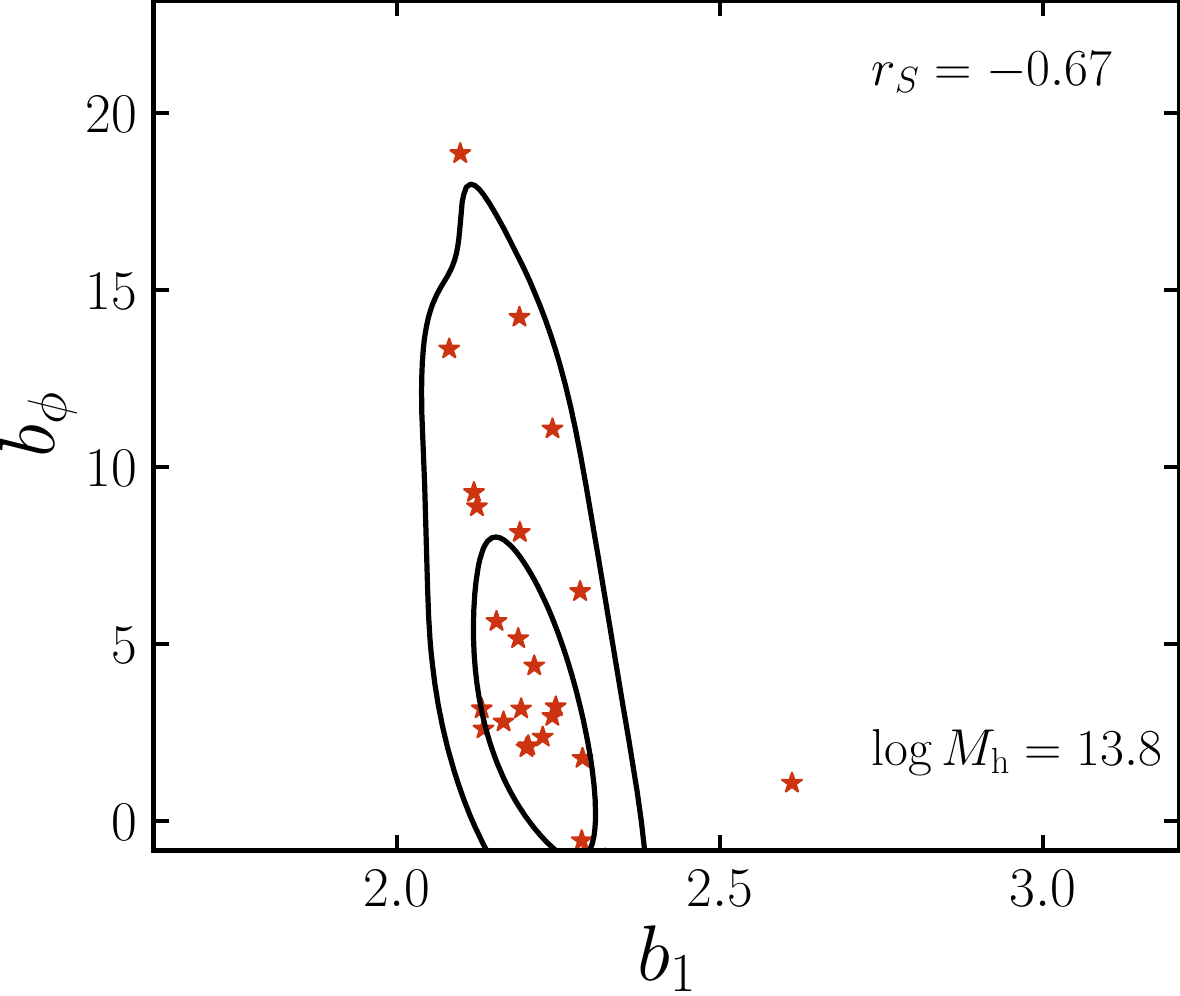}
    \includegraphics[width=.195\linewidth]{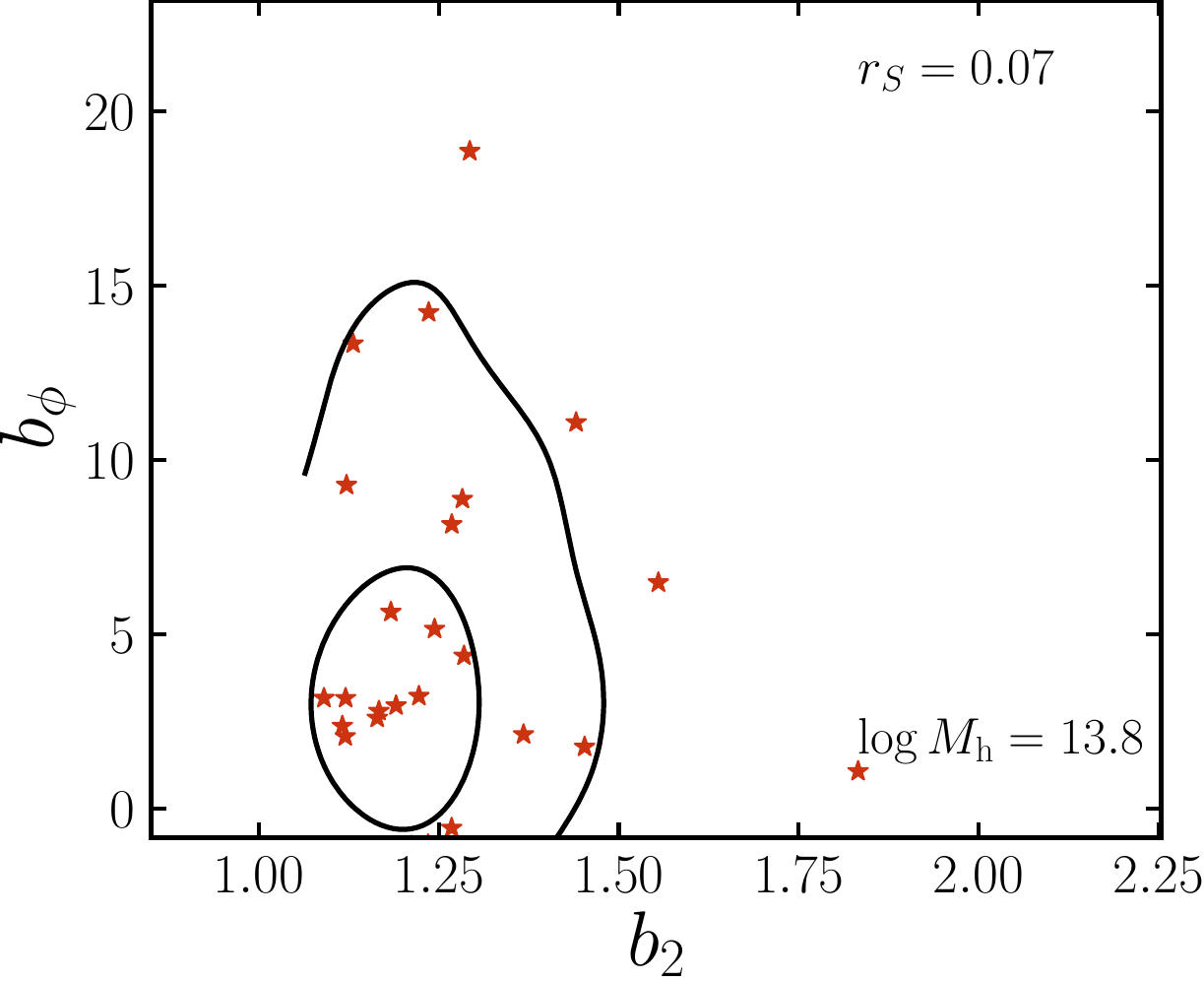}
    \includegraphics[width=.195\linewidth]{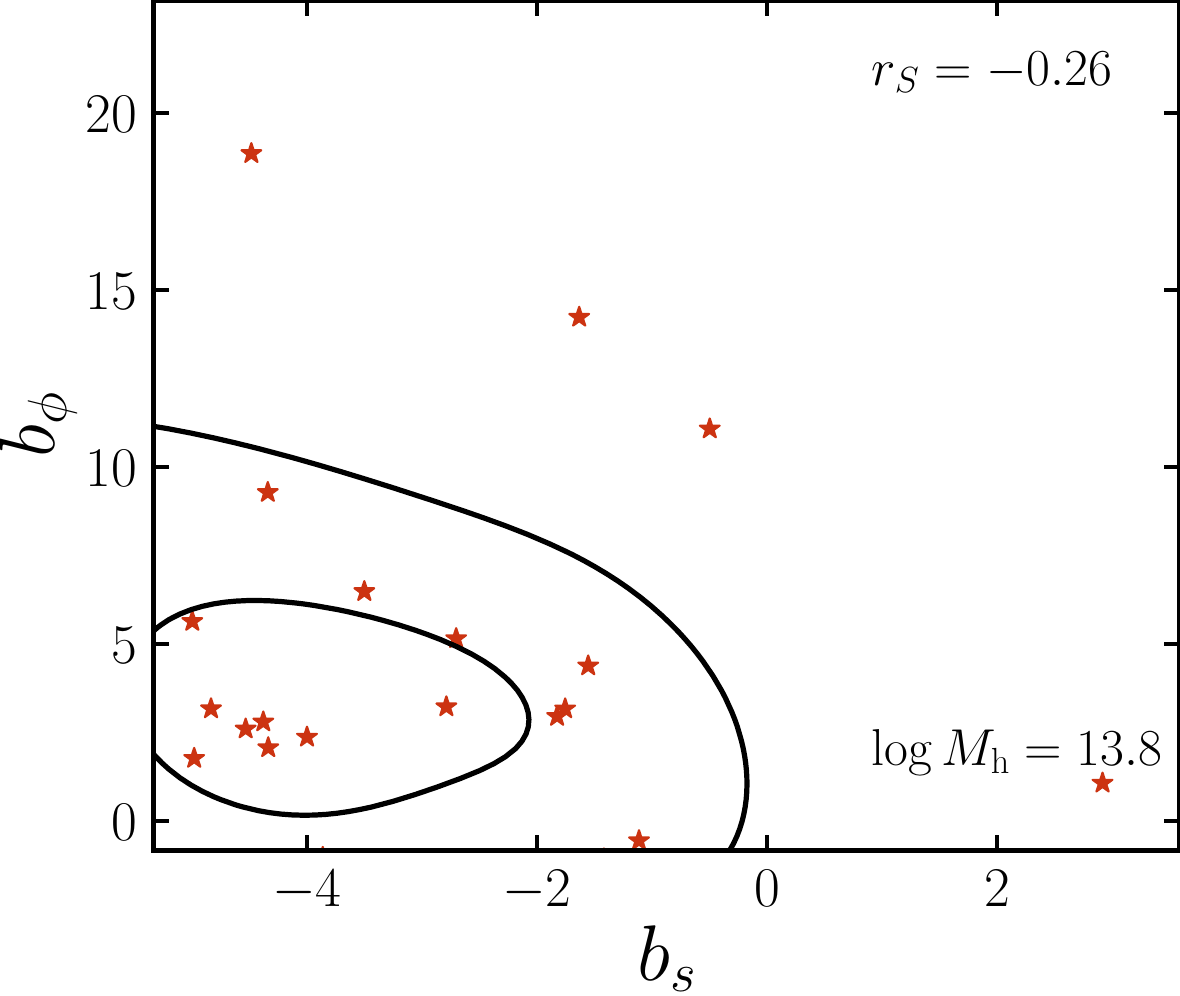}
    \includegraphics[width=.195\linewidth]{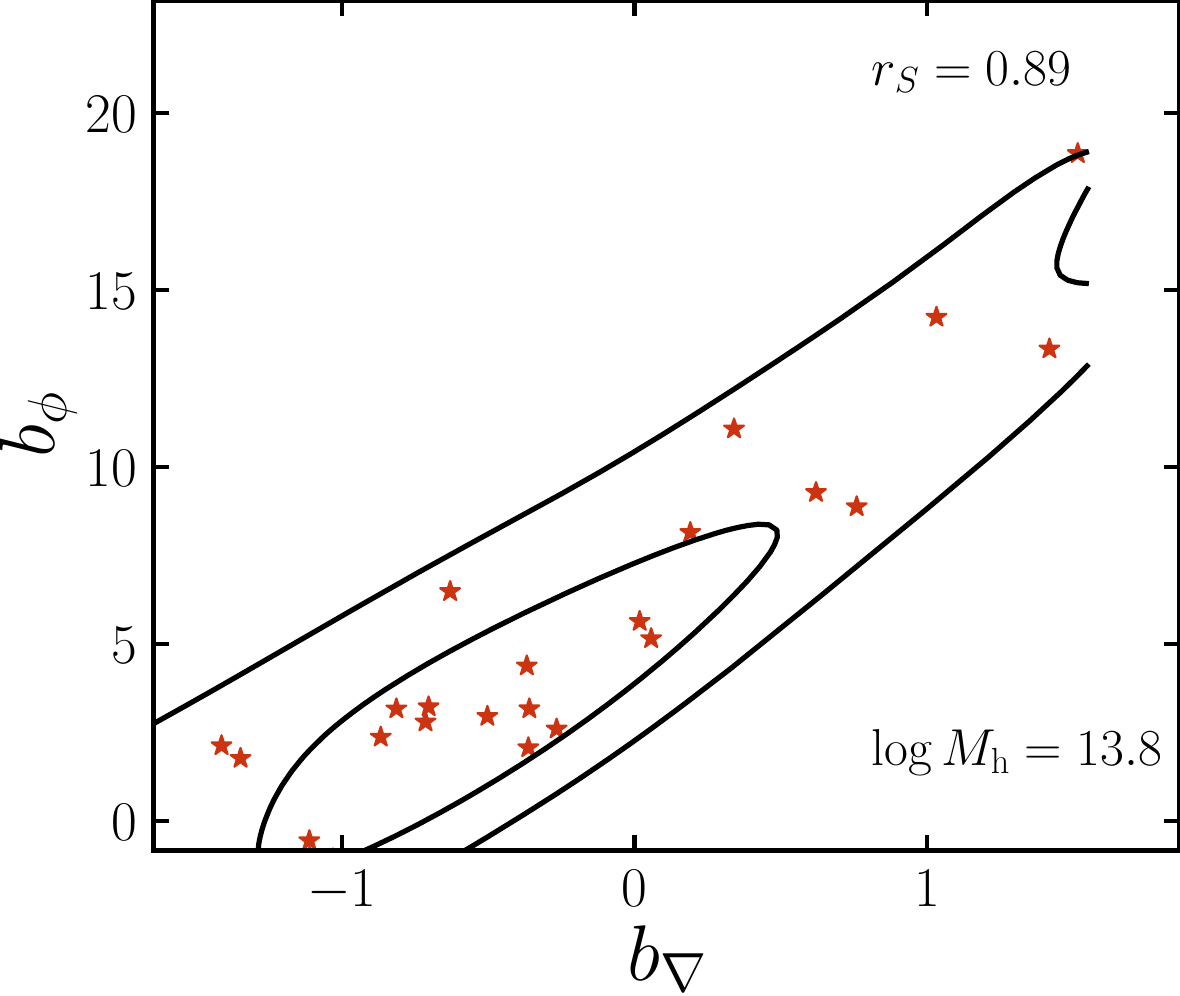}
    \includegraphics[width=.195\linewidth]{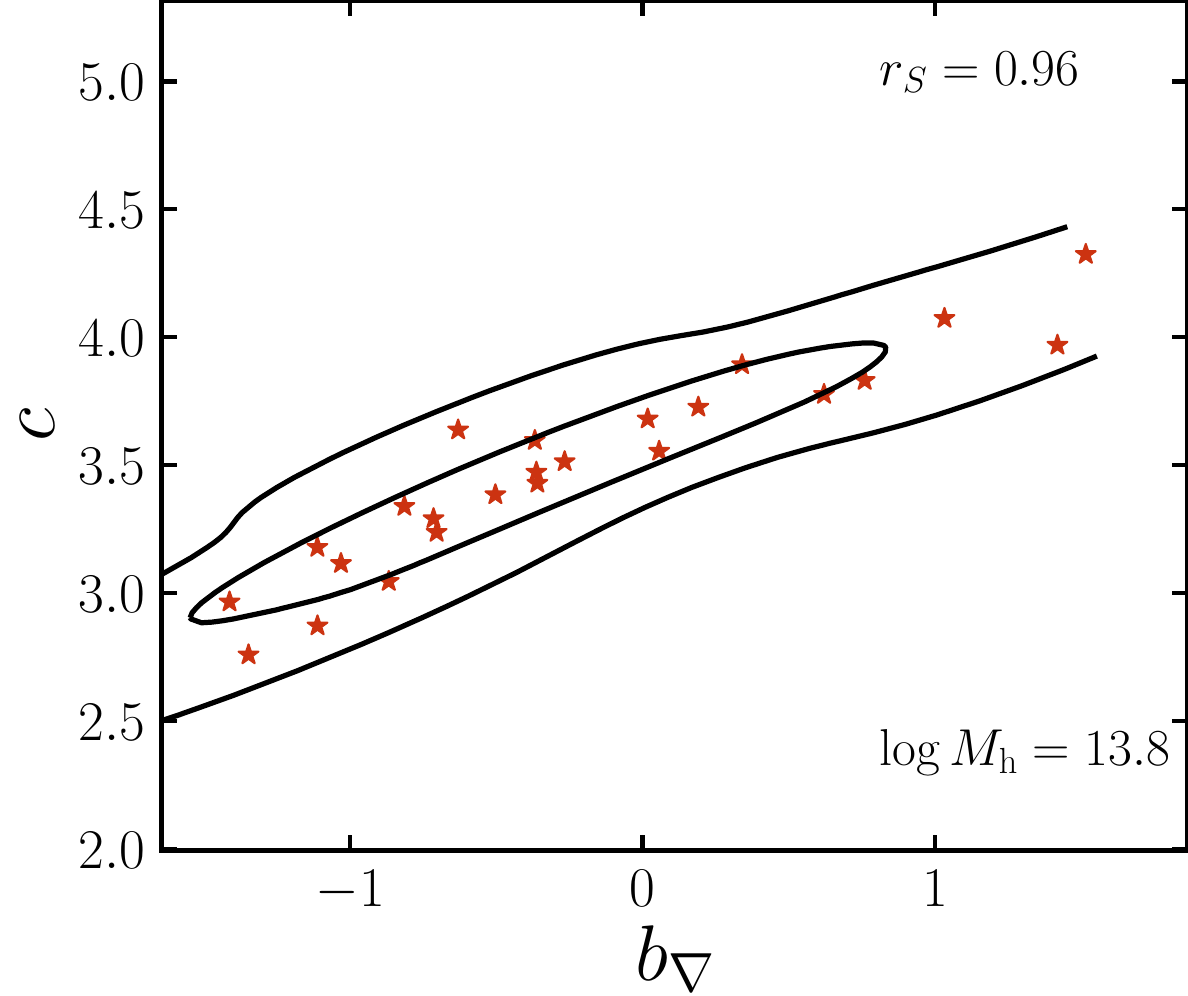}
    \caption{Correlation between $b_\phi$ and the 4 bias parameters at fixed halo mass, shown for two halo masses, $\log M_{\rm h} = 13.3, \ 13.8$, with contours indicating 68\% and 95\% percentile. A high Spearman correlation coefficient, $r_S$, means that if we know well the mass of the halo sample and the value of that parameter, we can pinpoint $b_\phi$ with a small amount of uncertainty. Looking at these plots, we see that consistently $b_1$ and $b_\nabla$ appear to be more highly correlated with $b_\phi$. We also find that concentration and $b_\nabla$ are very strongly correlated -- with Spearman coefficients around 0.95 for the mass range of interest. In other words, if one knows the concentration or more practically $b_\nabla$ for a given halo tracer, then identifying $b_\phi$ can be done with a small error bar. We note that in practice we rarely work directly with the halo field and often $b_\nabla$ absorbs counterterms as well as baryonic and astrophysical non-local effects.}
    \label{fig:halo_ind}
\end{figure*}

We show this in Fig.~\ref{fig:halo_ind} for two halo masses, $\log M_{\rm h} = 13.3, \ 13.8$, which are of interest to current and future surveys, with the contours indicating 68\% and 95\% percentile. The insight we get from studying these two mass bins is qualitatively similar to other mass bins, so we opt to only show those two. In each panel, we indicate the Spearman correlation, $r_S$ coefficient between the two parameters with positive values indicating a positive correlation and vice versa. Values of $r_S$ close to zero correspond to no correlation and $|r_S| = 1$ is maximal correlation. A high Spearman correlation coefficient between $b_\phi$ and some bias parameter $b_X$ means that if we know well the mass of the halo sample and the value of that parameter, we can pinpoint $b_\phi$ with a small amount of uncertainty. Looking at these plots, we see that consistently $b_1$ and $b_\nabla$ appear to be more highly correlated with $b_\phi$ (though in the case of $b_1$ it is only $\sim$65\%): in fact, the correlation between $b_\nabla$ and $b_\phi$ is so high that at fixed mass knowing one implies knowing the other with a high level of certainty. If one has direct access to a halo catalog, this is a very interesting and encouraging result.

To track down where this dependence arises from, we also study the correlation between concentration and $b_\nabla$ as the last panel in Fig.~\ref{fig:halo_ind}. Given the strong response of $b_\phi$ to both $b_\nabla$ and concentration, we find that concentration and $b_\nabla$ are very strongly correlated -- with Spearman coefficients around 0.95 for the mass range of interest. To our knowledge, this tight correlation between concentration and $b_\nabla$ has not been pointed out previously. The connection between the two can be gleaned from the physical interpretation of $b_\nabla$ as the parameter specifying the response of the halo field to the curvature of the matter field. In other words, if one knows the concentration or more practically $b_\nabla$ for a given halo tracer, then identifying $b_\phi$ can be done with a small error bar. We note that in practice we rarely work directly with the halo field and often $b_\nabla$ absorbs counterterms as well as baryonic and astrophysical non-local effects.

\subsubsection{Reducing the uncertainty}

\begin{figure*}
    \centering
    \includegraphics[width=.24\linewidth]{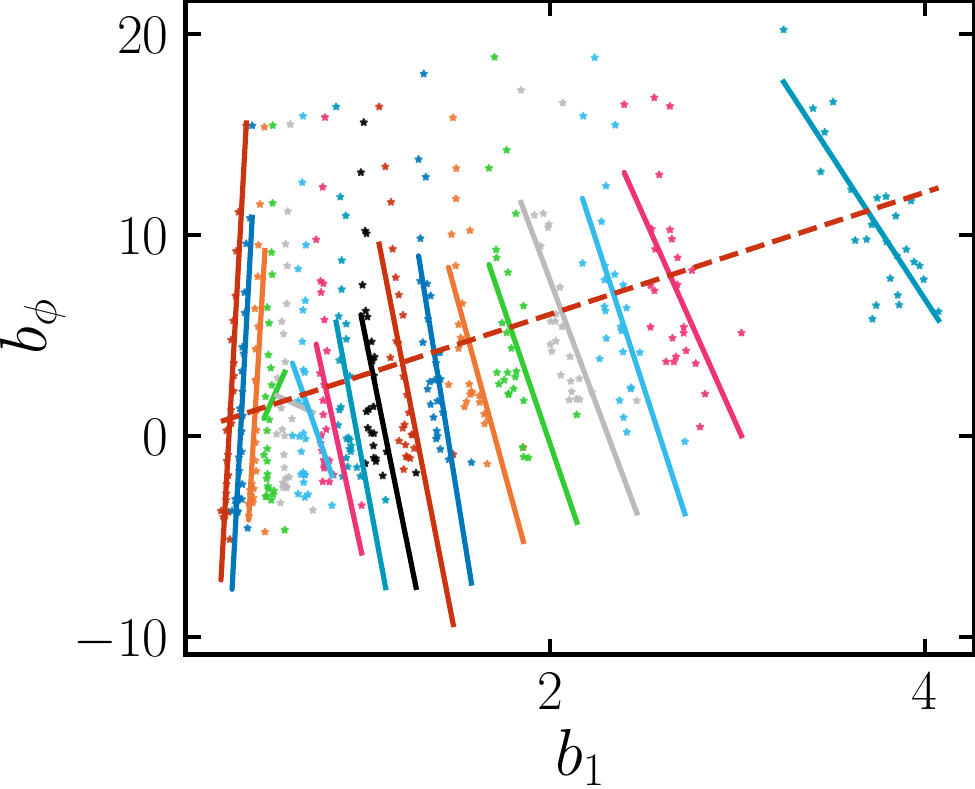}
    \includegraphics[width=.24\linewidth]{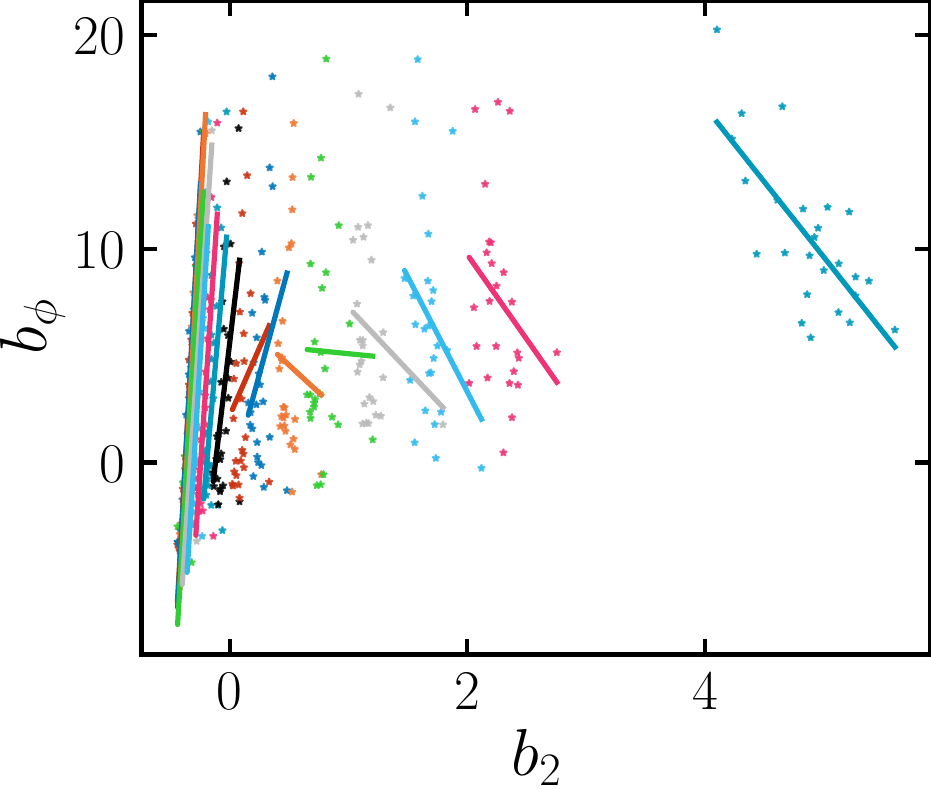}
    \includegraphics[width=.24\linewidth]{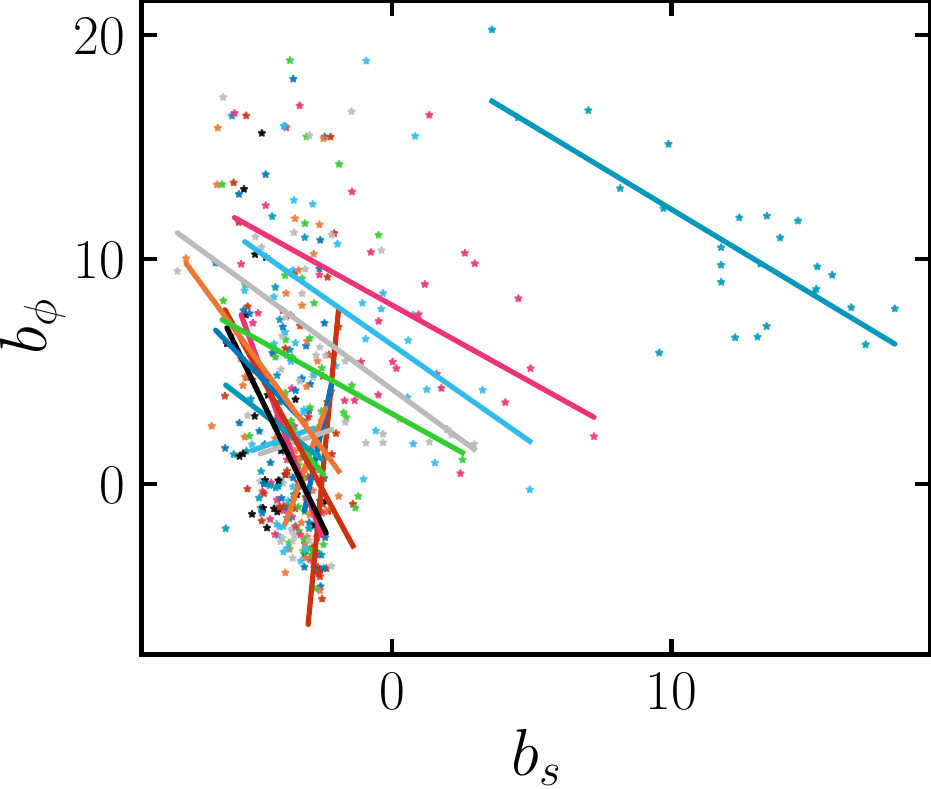}
    \includegraphics[width=.24\linewidth]{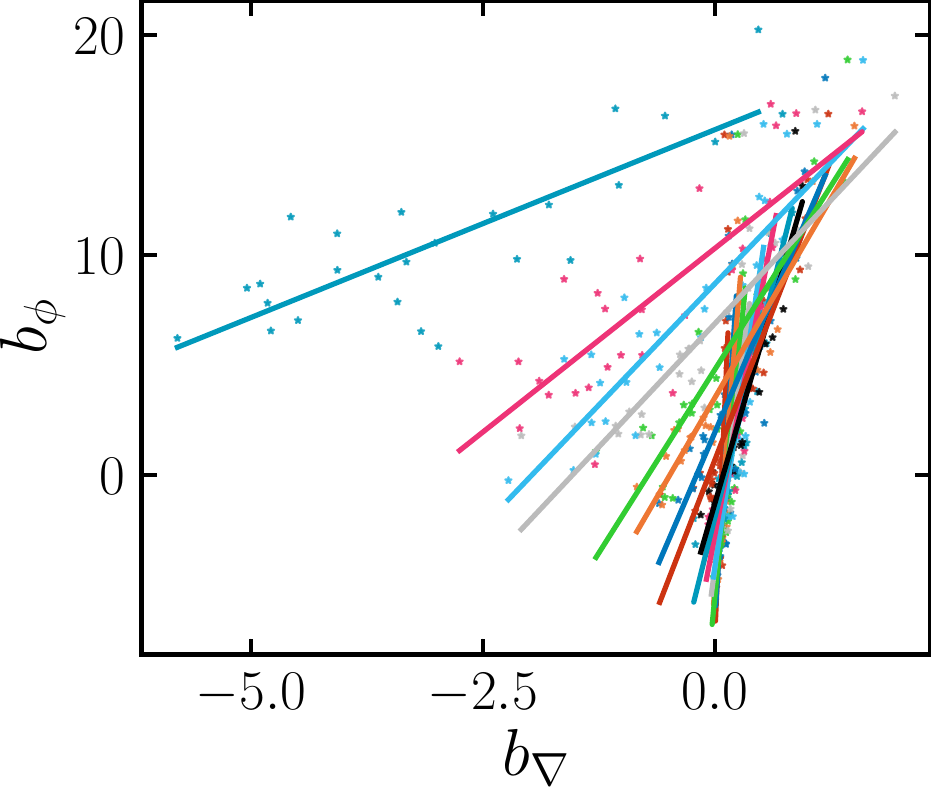}
    \caption{Scatter plot of the values of the Lagrangian bias parameters and $b_\phi$, where we have split the halos into 17 mass bins and 25 concentration bins. Halos belonging to the same mass bin are demarcated with the same color. The colored lines show the best-fit solution in the case of a single-parameter model, i.e. $b_\phi = \beta_0 + \beta_X b_X$, where $X$ stands for one of the four Lagrangian bias parameters. For the higher-mass bin, the scatter in $b_X$ for all four parameters is very large and thus, measuring $b_X$ (in addition to the halo mass) yields an accurate estimate of $b_\phi$, as can be seen in Table~\ref{tab:halo_improv}, where for the highest-mass bin, all four single-parameter models lead to a large reduction on the uncertainty of $b_\phi$. }
    \label{fig:halo_all}
\end{figure*}

\begin{table}
\begin{center}
\begin{tabular}{c c c c c c}
 \hline\hline
Halo mass & ${\rm All} \ b's$ & ${\rm Only} \ b_1$ & ${\rm Only} \ b_2$ & ${\rm Only} \ b_s$ & ${\rm Only} \ b_\nabla$ \\ [0.5ex]
 \hline
$logM_{\rm h} = 12.6$ & 76.0\% & 61.6\% & 74.3\% & 21.9\% & 31.4\% \\ [1ex]
$logM_{\rm h} = 12.7$ & 74.4\% & 37.4\% & 70.5\% & 4.3\% & 30.6\% \\ [1ex]
$logM_{\rm h} = 12.8$ & 79.6\% & 21.8\% & 73.3\% & 2.9\% & 37.7\% \\ [1ex]
$logM_{\rm h} = 12.9$ & 70.7\% & 0.4\% & 54.2\% & 0.6\% & 40.5\% \\ [1ex]
$logM_{\rm h} = 13.0$ & 75.4\% & 0.0\% & 59.1\% & 0.1\% & 43.3\% \\ [1ex]
$logM_{\rm h} = 13.1$ & 68.1\% & 1.7\% & 42.5\% & 0.1\% & 54.8\% \\ [1ex]
$logM_{\rm h} = 13.2$ & 73.4\% & 7.2\% & 33.3\% & 12.7\% & 67.7\% \\ [1ex]
$logM_{\rm h} = 13.3$ & 79.8\% & 14.8\% & 19.4\% & 1.3\% & 59.7\% \\ [1ex]
$logM_{\rm h} = 13.4$ & 76.4\% & 16.0\% & 15.7\% & 10.0\% & 70.6\% \\ [1ex]
$logM_{\rm h} = 13.5$ & 79.8\% & 27.3\% & 1.3\% & 13.3\% & 69.0\% \\ [1ex]
$logM_{\rm h} = 13.6$ & 70.9\% & 26.6\% & 4.7\% & 2.5\% & 60.0\% \\ [1ex]
$logM_{\rm h} = 13.7$ & 75.8\% & 23.6\% & 0.4\% & 15.7\% & 70.7\% \\ [1ex]
$logM_{\rm h} = 13.8$ & 71.6\% & 13.2\% & 0.0\% & 3.4\% & 55.8\% \\ [1ex]
$logM_{\rm h} = 13.9$ & 66.4\% & 24.7\% & 2.0\% & 14.4\% & 56.4\% \\ [1ex]
$logM_{\rm h} = 14.0$ & 58.1\% & 22.8\% & 4.3\% & 9.4\% & 51.1\% \\ [1ex]
$logM_{\rm h} = 14.1$ & 45.9\% & 18.6\% & 4.1\% & 11.1\% & 39.5\% \\ [1ex]
$logM_{\rm h} = 14.8$ & 47.2\% & 41.2\% & 32.5\% & 34.9\% & 40.2\% \\ [1ex]
All halos & 35.3\% & 10.1\% & 10.6\% & 2.6\% & 0.1\% \\ [1ex]
 \hline
 \hline
\end{tabular}
\end{center}
\caption{
At small halo masses, the performance of the $b_1$ and $b_2$ single-parameter models is very high, with the $b_2$ one being able to account for $\sim$70\% of the scatter. In the higher mass range, relevant to current surveys, we see that the single-parameter model that provides the best results is the one based on $b_\nabla$, which routinely explains well above 60\% of the $b_\phi$ uncertainty. The full-parameter model always does better than any of the single-parameter models. Specifically, for the halo masses most relevant for large-scale surveys, i.e. $\log M_{\rm h}$ between 13 and 14, the reduction of uncertainty on $b_\phi$ and hence $\fnlloc$ is around 75\%. While we are interested in studying the performance of this model at fixed halo mass, we also show the result for all halos for completeness, in the last line of Table~\ref{tab:halo_improv}. The reduction in uncertainty is only 35\% in the full model and none of the single-parameter models can explain more than 10\% of that uncertainty on its own.}
\label{tab:halo_improv}
\end{table}

We study by how much we can reduce the uncertainty of $b_\phi$ for the different halo mass bins next. Intuitively, we are trying to find what bias parameter or combination of bias parameters can capture best the scatter around the mean $b_\phi$ at fixed halo mass. While concentration is not a perfect proxy of $b_\phi$, it does correlate strongly with it, and varying it does allow us to cover the full range of $b_\phi$ values that a given halo sample can take. For this reason, we split the halos in each mass bin into 25 concentration bins and measure the bias parameters, $b_1$, $b_2$, $b_s$, $b_\nabla$, and $b_\phi$, for each mass-concentration bin\footnote{The goal of the mass-concentration binning is simply to capture the full range of values $b_\phi$ can take and look for trends with the Lagrangian bias parameters.}. Thus, assuming that the mass of the halo sample is known well, we proceed to fit $b_\phi$ as a function of the Lagrangian bias parameters for each mass bin. Specifically, we look for the linear coefficients, $\beta$, multiplying the biases that minimize the error on $b_\phi$, i.e.
\begin{equation}
    Y = \beta X ,
\end{equation}
where $Y^i = b_\phi^i$ and $X^i = [1, b_1^i, b_2^i, b_s^i, b_\nabla^i]$, with $i$ running from 0 to $N_{\rm conc} - 1 = 24$. The solution to this equation is:
\begin{equation}
    \beta = (X^T X)^{-1} X^T Y ,
    \label{eq:beta}
\end{equation}
and we note that $\beta_0$ here is the $Y$ offset. Technically, each entry $i$ should be weighted differently to reflect the concentration-based sampling of $b_\phi$, which is not necessarily properly weighting the different regions of $b_\phi$ space. However, we have tested that assigning coherently perturbed weights does not substantially alter our results. 

Our metric for how well we capture the scatter is as follows: at fixed halo mass, we can calculate the uncertainty on $b_\phi$ as ${\rm Std}[b_\phi^{\rm data}]$. This is the price we need to pay on our uncertainty on $\fnlloc$, since 
\begin{equation}
    \frac{\sigma[\fnlloc]}{|\fnlloc|} \propto \frac{\sigma[b_\phi]}{b_\phi} .
\end{equation}
We would now like our model, $b_\phi^{\rm model} \equiv \beta X$, to be as close as possible to $b_\phi^{\rm data}$, i.e. we want to minimize the scatter in $\Delta b_\phi \equiv b_\phi^{\rm data} - b_\phi^{\rm model}$ (this is in fact the solution we find in Eq.~\ref{eq:beta}). We can quantify how well we have done through an improvement percentage, quantified as:
\begin{equation}
    I [\%] \equiv {\rm} 100 \times \left(1 - \frac{{\rm Std}[\Delta b_\phi]}{{\rm Std}[b_\phi^{\rm data}]} \right) .
\label{eq:improv}
\end{equation}
We will focus on two types of models: those in which we make a fit with all Lagrangian parameters present, and those in which we only fit $b_\phi$ with one of the Lagrangian parameters at a time. These results are shown in Fig.~\ref{fig:halo_all} and Table~\ref{tab:halo_improv} at $z = 0.5$. Our findings are qualitatively similar for other redshifts.

In Fig.~\ref{fig:halo_all}, we show a scatter plot of the values of the Lagrangian bias parameters and $b_\phi$, where each point corresponds to a concentration-mass bin (we split the halos into 17 mass bins and 25 concentration bins). Halos belonging to the same mass bin are demarcated with the same color, as our analysis assumes that we are working with a halo sample at fixed halo mass (e.g. identified through a cluster finding algorithm). The colored lines show the best-fit solution in the case of a single-parameter model, i.e. $b_\phi = \beta_0 + \beta_X b_X$, where $X$ stands for one of the four Lagrangian bias parameters. The intuition in case of the single-parameter model is as follows: in cases where the scatter of the points is small in $b_X$, we are not able to pinpoint the value of $b_\phi$ well, as knowing $b_X$ does not give us any additional information. As an example, we see that for the higher-mass bin, the scatter in $b_X$ for all four parameters is very large and thus, measuring $b_X$ (in addition to the halo mass) yields an accurate estimate of $b_\phi$, as can be seen in Table~\ref{tab:halo_improv}, where for the highest-mass bin, all four single-parameter models lead to a large reduction on the uncertainty of $b_\phi$. 

Similarly, on the small-mass end, we see from the table that the performance of the $b_1$ and $b_2$ single-parameter models is very high, with the $b_2$ one being able to account for $\sim$70\% of the scatter. Looking at Fig.~\ref{fig:halo_all}, we see that the variation of $b_1$ and $b_2$ is small for these low-mass samples, but it is very tight. Specifically, it appears that the scatter points $b_1$-$b_\phi$ and $b_2$-$b_\phi$ form a very thin straight line that is well captured by a single linear relationship. So for these very low-mass halos, $\log M_{\rm h} \lesssim 13$, if the mass and $b_2$ are known (notice that $b_2$ performs better than $b_1$), then $b_\phi$ can be predicted with a high accuracy. A number of current and future surveys, however, will focus on the higher-mass regime, $\log M_{\rm h} \gtrsim 13$. In this mass range, we see that the single-parameter model that provides the best results is the one based on $b_\nabla$, which routinely explains well above 60\% of the $b_\phi$ uncertainty. We link this to our previous finding from Fig.~\ref{fig:halo_ind} that $b_\nabla$ is an excellent proxy for concentration and as a result tightly correlated with $b_\phi$. 

In theory, using all bias parameters ought to always yield the best performance, surpassing or doing as well as fewer-parameter models, as it is able to define a hyperplane that describes the data more accurately. This is indeed what we see in the table: the full-parameter model always does better than any of the single-parameter models. Specifically, for the halo masses most relevant for large-scale surveys, i.e. $\log M_{\rm h}$ between 13 and 14, the reduction of uncertainty on $b_\phi$ and hence $\fnlloc$ is around 75\%. While we are interested in studying the performance of this model at fixed halo mass, we also show the result for all halos for completeness, in the last line of Table~\ref{tab:halo_improv}. Interestingly, the reduction in uncertainty is only 35\% in the full model and none of the single-parameter models can explain more than 10\% of that uncertainty on its own. We surmise that this is the result of mixing halo samples that have different responses to $b_\phi$ and the Lagrangian bias parameters. In addition, not imposing a fixed halo mass introduces degeneracies between the Lagrangian bias parameters and halo mass.


\subsection{Galaxies}

\begin{figure*}
    \centering
    \includegraphics[width=.24\linewidth]{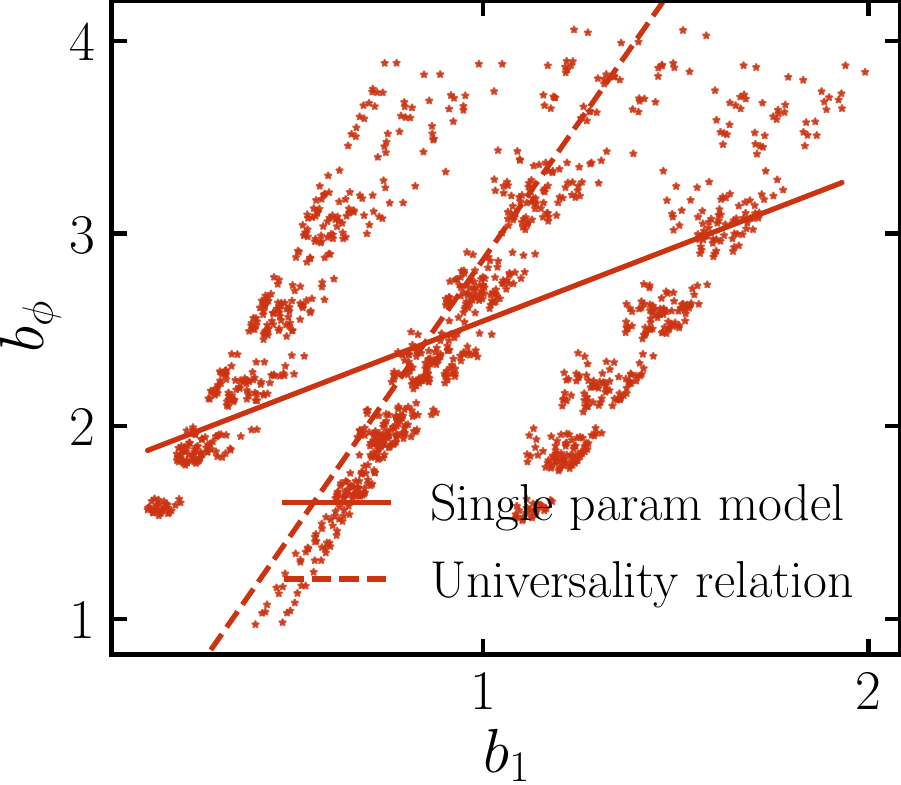}
    \includegraphics[width=.24\linewidth]{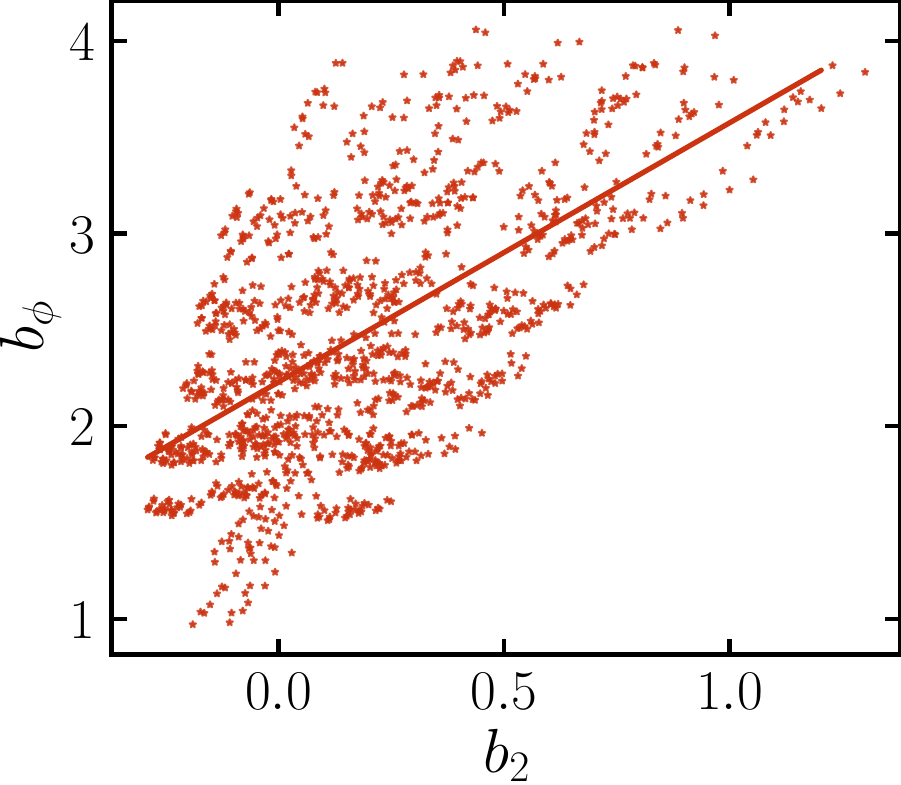}
    \includegraphics[width=.24\linewidth]{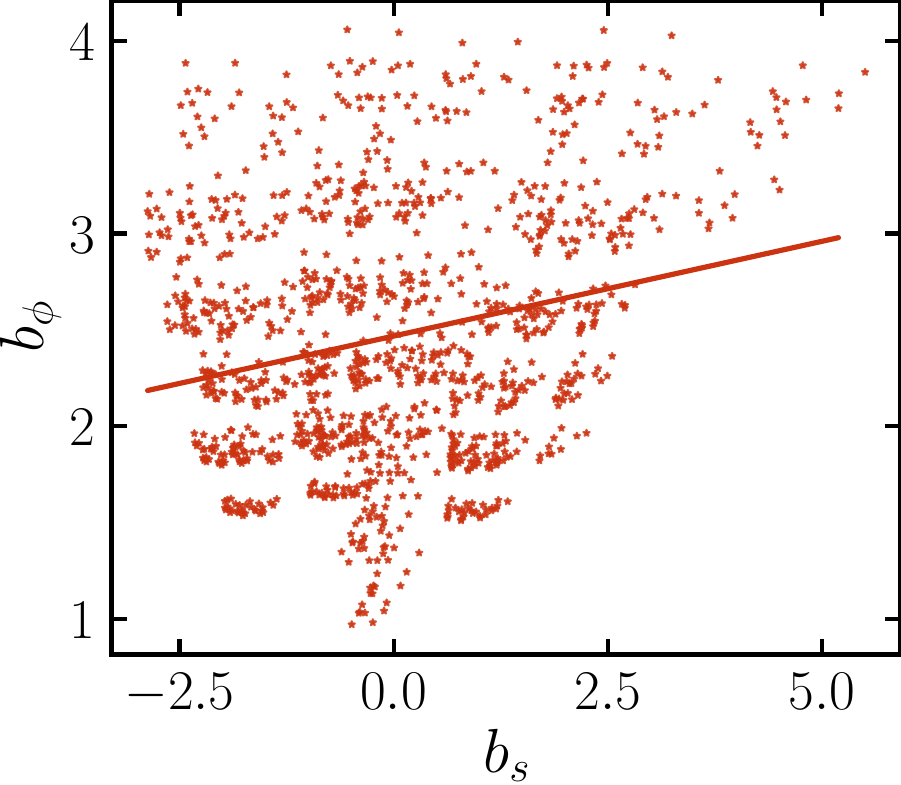}
    \includegraphics[width=.24\linewidth]{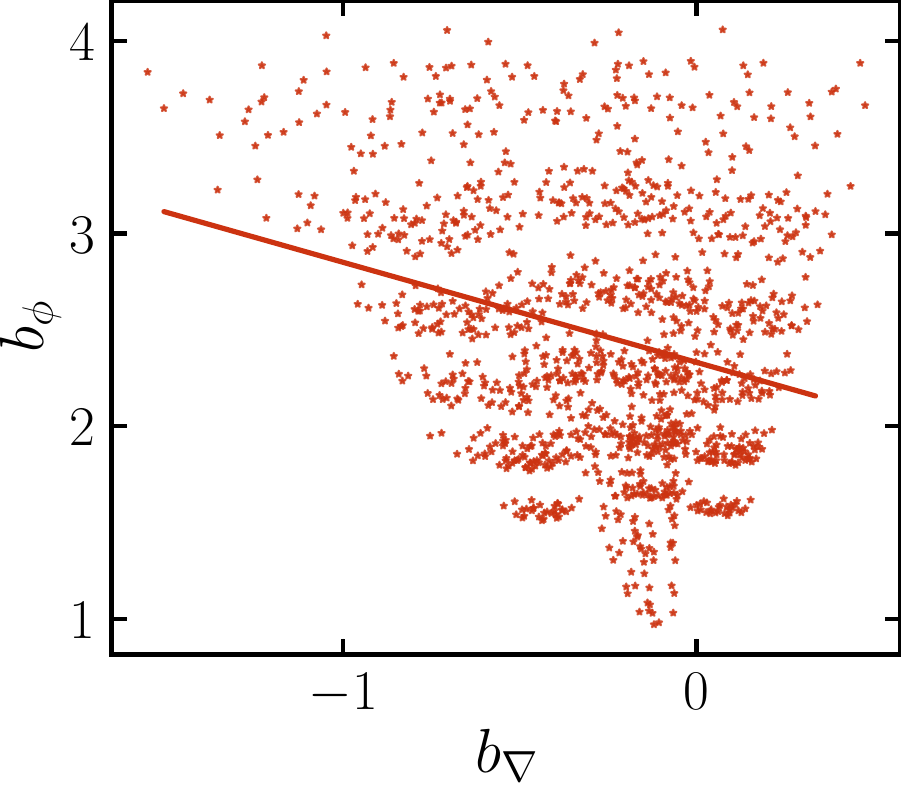}
    \caption{Scatter plots of the $b_X$-$b_\phi$ plane for the LRGs at $z = 0.5$. The three islands visible in the first panel, $b_1$-$b_\phi$, correspond to the low, mid and high-environmental assembly biases cases from left to right, respectively. Single values of $b_1$ can correspond to a large range of possibilities for the environmental assembly bias parameter. The dashed line demarcates the universality relation between $b_1$ and $b_\phi$ for each tracer (with $c = 0.85$, see Eq.~\ref{eq:univ}), which passes right through the no environmental assembly bias island in all cases. The solid line, as in the case of Fig.~\ref{fig:halo_all}, shows the best-fit single-parameter model, $b_\phi = \beta_0 +\beta b_X$ for each of the four Lagrangian bias parameters. 
}
    \label{fig:gal_lrg1}
\end{figure*}

\begin{figure*}
    \centering
    \includegraphics[width=.24\linewidth]{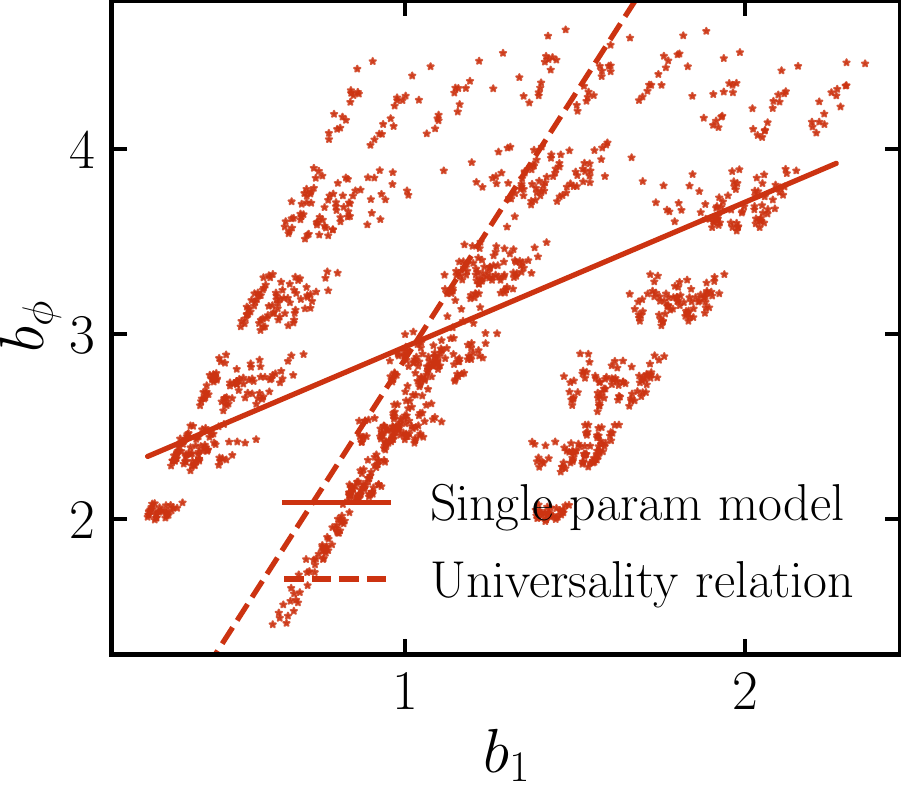}
    \includegraphics[width=.24\linewidth]{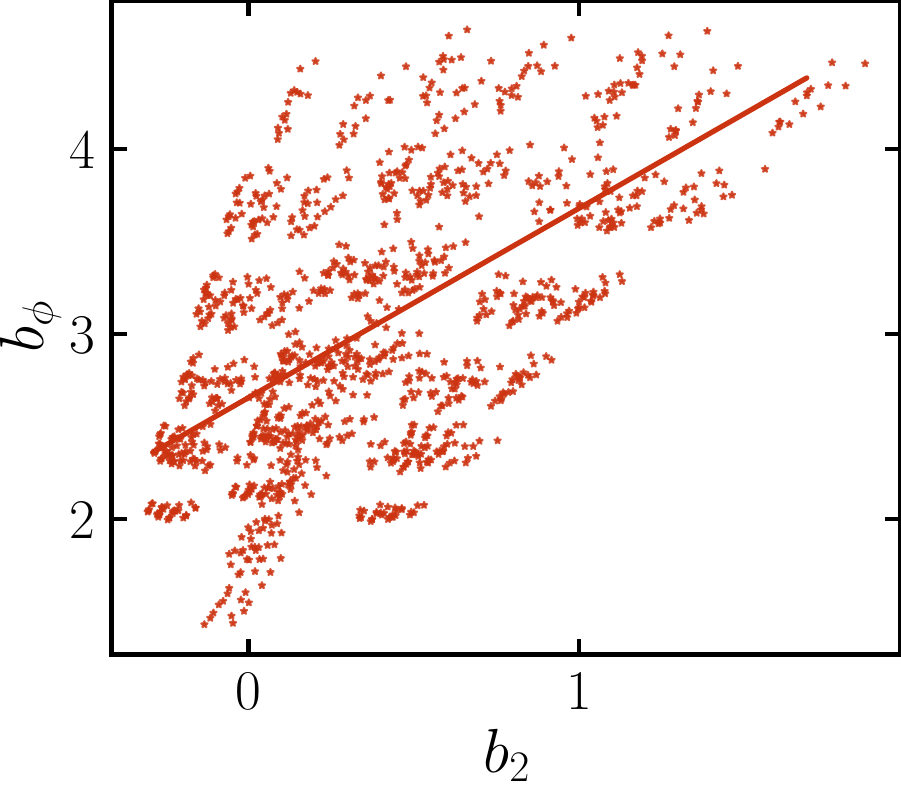}
    \includegraphics[width=.24\linewidth]{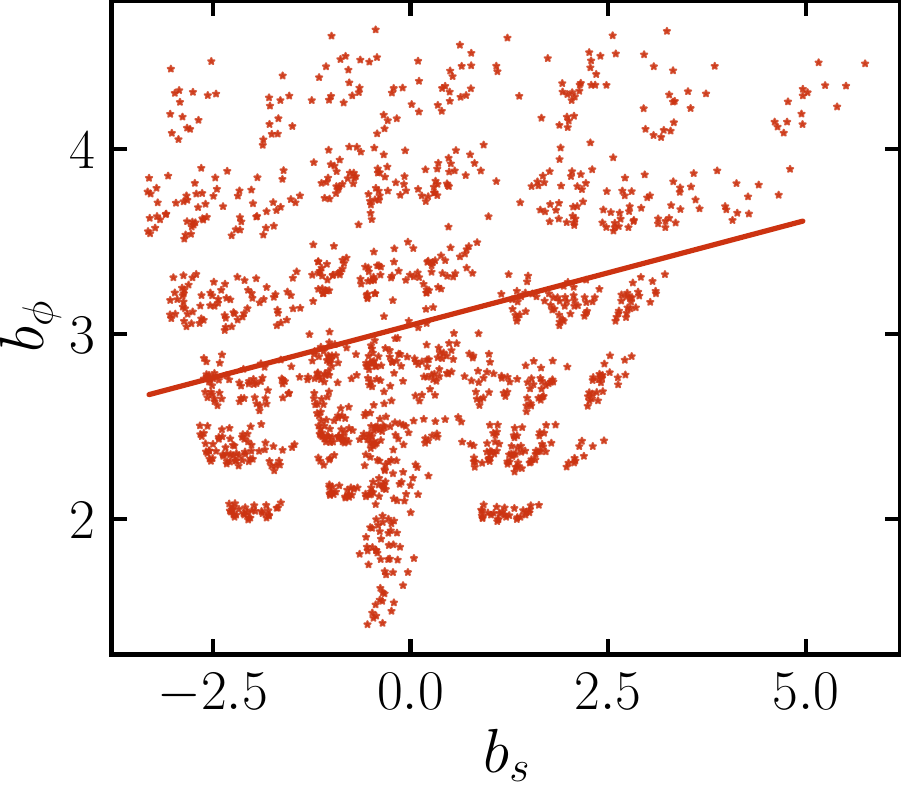}
    \includegraphics[width=.24\linewidth]{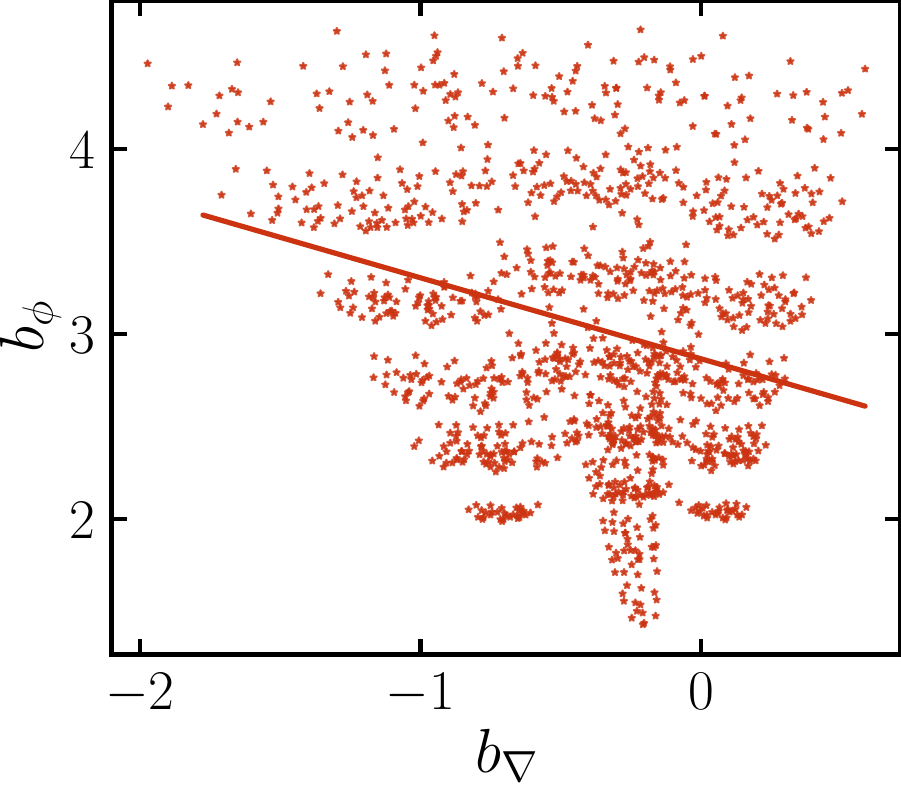}
    \caption{As Fig.~\ref{fig:gal_lrg1}, but for the second LRG sample at $z = 0.8$. The two LRG samples appear to be qualitatively very similar with only small differences: at $z = 0.8$, it seems like all bias parameters cover a slightly broader range of values compared with the LRGs at $z = 0.5$.}
    \label{fig:gal_lrg2}
\end{figure*}

\begin{figure*}
    \centering
    \includegraphics[width=.24\linewidth]{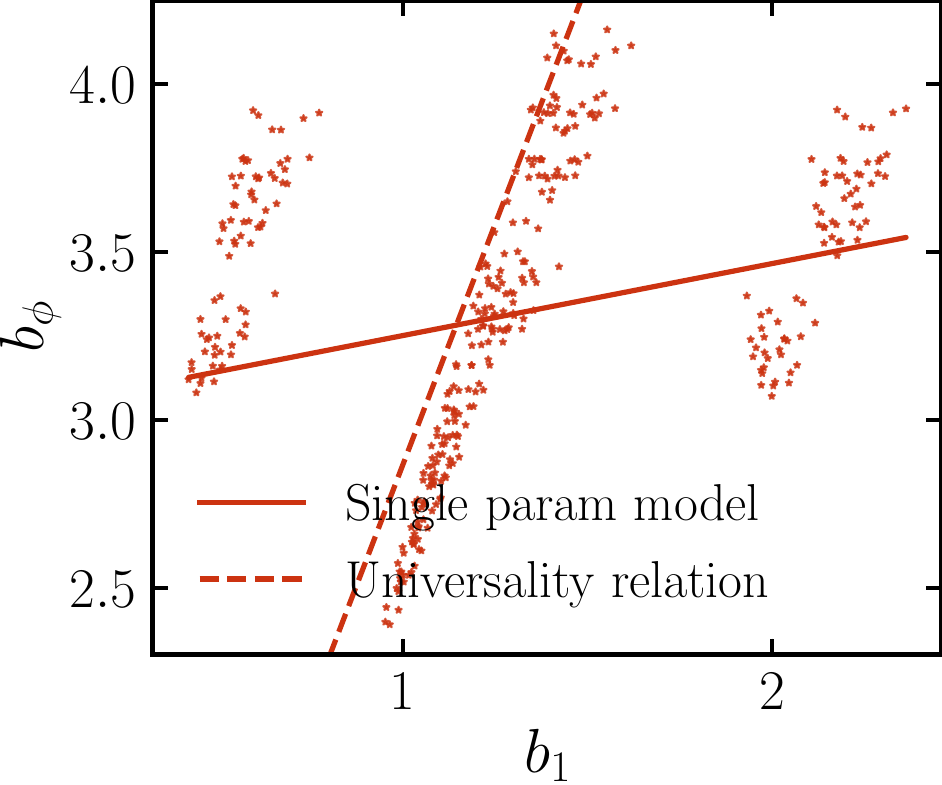}
    \includegraphics[width=.24\linewidth]{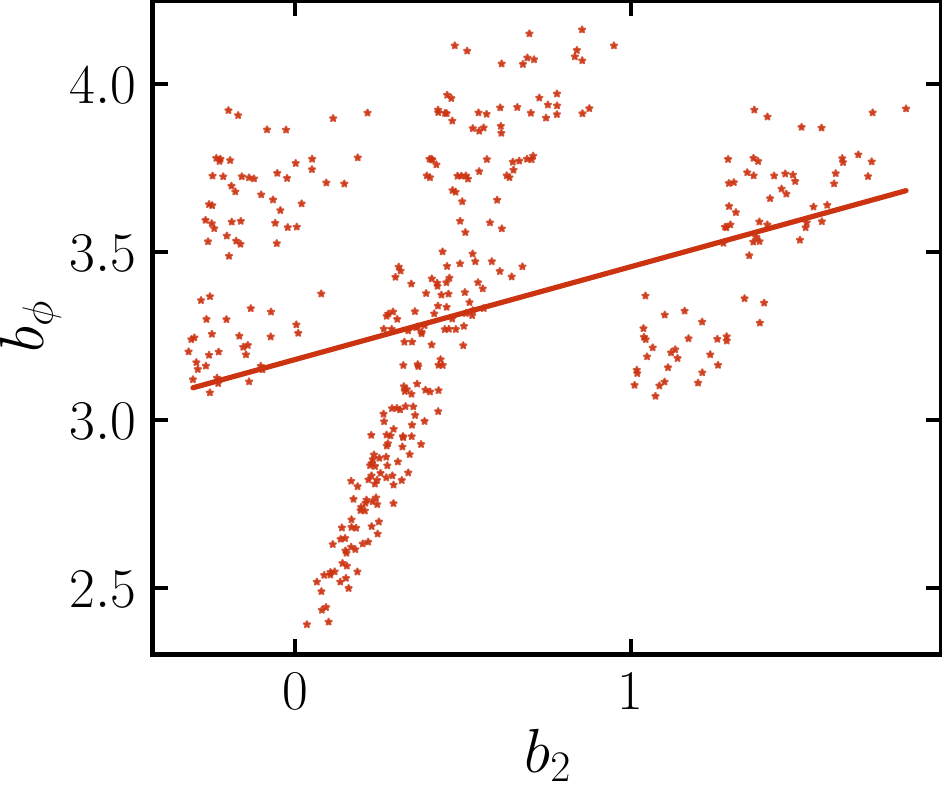}
    \includegraphics[width=.24\linewidth]{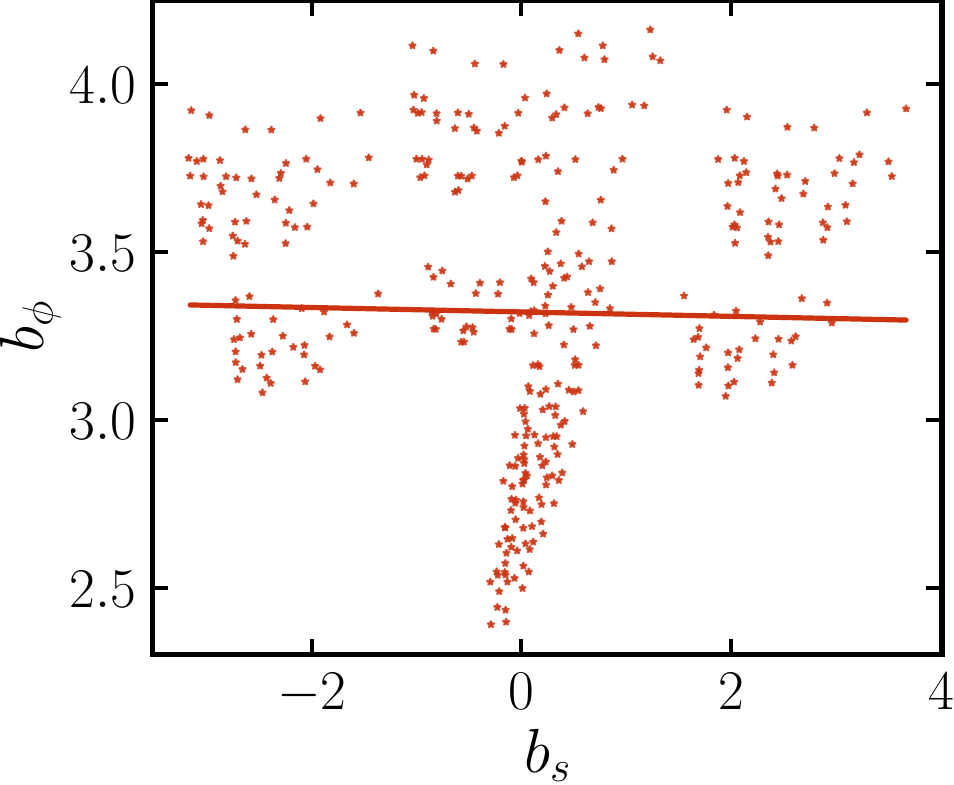}
    \includegraphics[width=.24\linewidth]{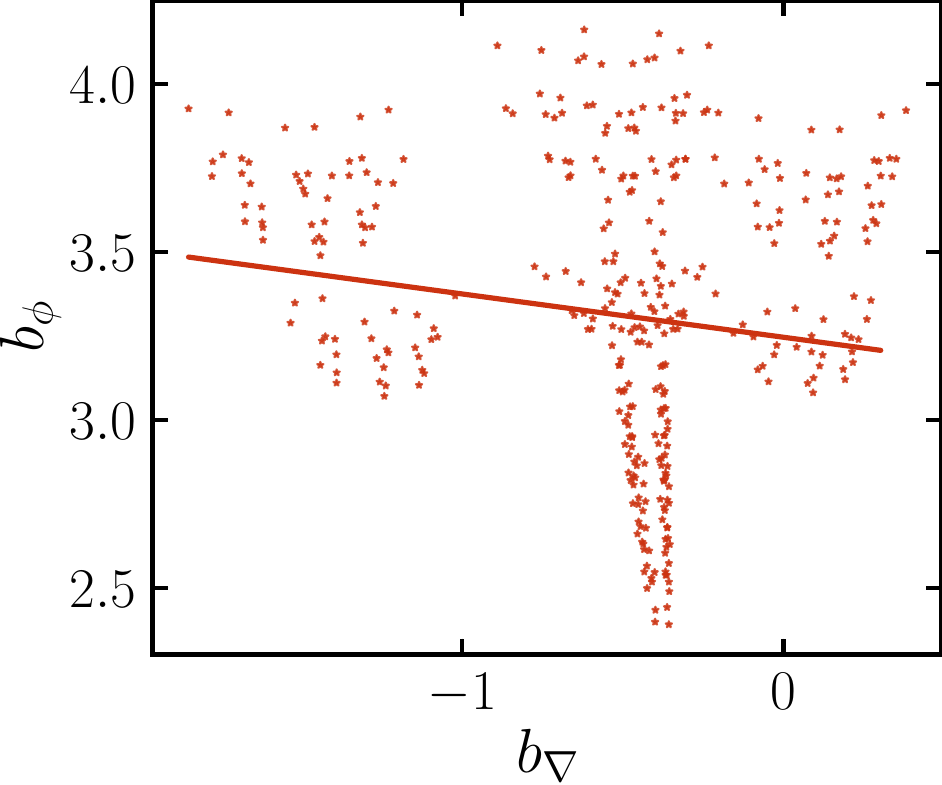}
    \caption{As Fig.~\ref{fig:gal_lrg1}, but for the second QSO sample at $z = 1.4$. The three environment islands are more separated compared with the LRG samples. This is likely the case, as at higher redshifts and for more biased tracers, selecting halos with a large or small number of neighbors as done for large and small values of $B_c$ and $B_s$ pinpoints the environment of a halo. Thus, in this case, knowing $b_1$ can break the degeneracy with environmental assembly bias sufficiently well. In addition, we note that $b_\phi$ does not appear to vary as much compared with the LRG samples.}
    \label{fig:gal_qso}
\end{figure*}

\begin{table}
\begin{center}
\begin{tabular}{c c c c c c}
 \hline\hline
Tracer, redshift & ${\rm All} \ b's$ & ${\rm Only} \ b_1$ & ${\rm Only} \ b_2$ & ${\rm Only} \ b_s$ & ${\rm Only} \ b_\nabla$ \\ [0.5ex]
 \hline
${\rm LRG}, \ z = 0.5$ & 81.6\% & 13.1\% & 22.3\% & 2.7\% & 2.8\% \\ [1ex]
${\rm LRG}, \ z = 0.8$ & 82.7\% & 10.7\% & 17.9\% & 1.9\% & 2.2\% \\ [1ex]
${\rm QSO}, \ z = 1.4$ & 81.8\% & 4.9\% & 6.8\% & 0.0\% & 1.2\% \\ [1ex]
 \hline
 \hline
\end{tabular}
\end{center}
\caption{Improvement percentages for the three tracer samples. While none of the single-parameter models reduces the scatter on $b_\phi$ by more than $\sim$20\% for the LRGs and $\sim$7\% for the QSOs, when combined together, the scatter is reduced by over $\gtrsim$80\% for all three tracers.}
\label{tab:gal_improv}
\end{table}

\begin{table*}
\begin{center}
\begin{tabular}{c c c c c c c c}
 \hline\hline
Tracer, redshift & $\beta_0$ & $b_1$ & $b_2$ & $b_s$ & $b_\nabla$ & ${\rm Std}[b_\phi^{\rm data}]$ & ${\rm Std}[\Delta b_\phi]$ \\ [0.5ex]
 \hline
${\rm LRG}, \ z = 0.5$ & 1.18 $\pm$ 0.08 & 0.12 $\pm$ 0.08 & 4.87 $\pm$ 0.11 & -0.98 $\pm$ 0.04 & -1.03 $\pm$ 0.16 & 0.63 & 0.12 \\ [1ex]
${\rm LRG}, \ z = 0.8$ & 1.74 $\pm$ 0.08 & -0.30 $\pm$ 0.06 & 4.24 $\pm$ 0.07 & -0.75 $\pm$ 0.04 & 0.09 $\pm$ 0.12 & 0.74 & 0.12 \\ [1ex]
${\rm QSO}, \ z = 1.4$ & 3.38 $\pm$ 0.11 & -0.54 $\pm$ 0.13 & 3.62 $\pm$ 0.16 & -0.30 $\pm$ 0.02 & 2.01 $\pm$ 0.12 & 0.38 & 0.09 \\ [1ex]
 \hline
 \hline
\end{tabular}
\end{center}
\caption{Best-fit values for the $\beta$ coefficients multiplying the four bias parameters,  $b_X$, and overall offset, $\beta_0$. Using these along with the measured four bias parameters in a simple regression model allows us to predict $b_\phi$ and can serve as a prior in future PNG analysis. It is reassuring that the best fit $\beta$ coefficients for the two LRG samples are very similar, whereas the quasar sample yields slightly different values for the $\beta$'s.}
\label{tab:gal_beta}
\end{table*}

\subsubsection{Constructing fair galaxy samples}

Most potent for the search of PNG are large-volume spectroscopic surveys such as DESI, which aims to place $\sigma[\fnlloc] \approx 5$ bounds through the auto-correlation analysis of its quasar (QSO) sample and its luminous red galaxy (LRG) samples. Here, we study all three samples in the cubic box at fixed redshift snapshots: LRG at $z = 0.5$, LRG at $z = 0.8$, and QSO at $z = 1.4$, constructed via the AbacusHOD model (see Section~\ref{sec:hod}). We do not adjust our analysis to the volume of the survey, as the goal of our exploration is to understand the connection between $b_\phi$ and the Lagrangian bias parameters $b_X$ with minimal noise. We find that using a single simulation box, \texttt{AbacusSummit\_base\_c000\_ph000} along with the field-level bias measuring technique (see Section~\ref{sec:heft}) and the Separate Universe approach (see Section~\ref{sec:univ}) is sufficient for our purposes.

We have already learned that the halo properties that have a strong dependence on the value of $b_\phi$ are the halo mass and assembly bias parameters such as concentration and environment. For this reason, in designing the galaxy samples for this study, we adopted the best-fit parameters of the extended HOD analysis of Ref.~\citep{2024MNRAS.530..947Y}, exploring modifications to six of the model parameters, namely: $\log M_{\rm cut}$, $\log M_1$, the two main parameters controlling the halo masses entering the sample, $A_{\rm c}$, $A_{\rm s}$, the concentration assembly bias parameters, and $B_{\rm c}$, $B_{\rm s}$, the environment assembly bias parameters. Specifically, we vary the mass parameters within 0.2 dex of their best-fit values, the concentration ones between -2 and 2, and the environment ones between -0.5 and 0.5. We note that some of these variations appear to be ruled out by the tight constraints placed on the data, but at the same time, assembly bias ornamentations are currently not too well constrained.

We show scatter plots of the $b_X$-$b_\phi$ plane for the three galaxy tracer samples in Fig.~\ref{fig:gal_lrg1}, \ref{fig:gal_lrg2}, and \ref{fig:gal_qso}. As expected, the range of variation for both $b_\phi$ and the $b_X$ parameters is more narrow compared with the halos, but the scatter still appears to be large. We note that the three islands visible in the first panel, $b_1$-$b_\phi$, for all three samples, correspond to the low, mid and high-environmental assembly biases cases from left to right, respectively. As noted previously, unlike concentration, environment leads to shifts in the $x$ axis (rather than the $y$ axis) \citep[see e.g.,][]{2024PhRvD.109j3530H}. We see that single values of $b_1$ can correspond to a large range of possibilities for the environmental assembly bias parameter. This accounts for the largest uncertainty on $b_\phi$ that we see in the samples we have defined.


\subsubsection{Reducing the uncertainty}

Unlike the halo case where we adopted fixed-mass samples, in the case of galaxies we are mixing very different halo masses and very different concentrations, which erases a lot of the information that turned out to be helpful in pinpointing the value of $b_\phi$ for halos. In addition, the presence of satellites and stochastic sampling makes the connection between concentration and $b_\nabla$ more complex compared with the halo case. In practice, we know that concentration assembly bias does affect the galaxy occupation distribution in non-trivial ways: high-concentration halos are more likely to host centrals, whereas low-concentration halos are more likely to host satellites \citep[see e.g.,][]{2021MNRAS.508..698H}. We do hope that the mix of halo masses and properties of the galaxy host halos can be disentangled by measuring a few of the Lagrangian bias parameters simultaneously. This is demonstrated in this section for the three galaxy samples of interest: LRGs at $z = 0.5$ and 0.8 and QSOs at $z = 1.4$. 

As before, we fit a linear regression model to the measured values of $b_\phi$ and $b_X$ for each of the HOD samples of a given tracer sample, of the form $b_\phi = \beta_0 + \sum_{X}\beta_X b_X$, where $X$ is one of four Lagrangian bias parameters: $b_1$, $b_2$, $b_s$, $b_\nabla$. We also evaluate the improvement percentage (see Eq.~\ref{eq:improv}) relative to the case where we do not try to model $b_\phi$ and instead adopt some best-guess value and an empirically derived uncertainty on it, ${\rm Std[b_\phi^{\rm data}]}$, and also relative to single-parameter models of the form $b_\phi = \beta_0 + \beta_X b_X$. We show the best-fit lines for the single-parameter case as solid lines in Figs.~\ref{fig:gal_lrg1}, \ref{fig:gal_lrg2}, and \ref{fig:gal_qso}. The $b_1$-$b_\phi$ universality relation is drawn as a dashed line in the first panel. We see that it goes almost exactly through the center of the middle `island', suggesting that if little or no assembly bias is present in the sample, it would provide a very good approximation to the value of $b_\phi$. Visually, we see some stark similarities between the two LRG samples in terms of the relationship between the bias parameters and $b_\phi$ for the adopted HOD samples: similar trends to the mean $b_X$-$b_\phi$ relations, as well as the scatter of these relations. This consistency is reassuring in that it attests to the fact that we are observing true trends, as opposed to noise in the samples we have identified. While the QSO sample is different in some obvious ways (e.g., mean bias values and size fo the scatter), we do find that the overall structure of the $b_X$-$b_\phi$ spaces is largely identical.

Looking at the improvement percentages in Table~\ref{tab:gal_improv}, we see something surprising: while none of the single-parameter models reduces the scatter on $b_\phi$ by more than $\sim$20\% for the LRGs and $\sim$7\% for the QSOs, when combined together the scatter is reduced by over $\gtrsim$80\% for all three tracers. This shows that there is a hyperplane that connects $b_\phi$ and the Lagrangian bias parameters much more effectively than any of the $b_\phi$-$b_X$ cross-sections. Interestingly, out of the single-parameter models it is not the one based on $b_1$ or the one based on $b_\nabla$, as the halo case would have led us to surmise, that gives the largest reduction on uncertainty, but instead it is the one based on $b_2$. We explain this by noting that the mean halo mass for e.g. the LRG samples is around $\log M_{\rm h} = 13.2$ and as can be seen from Table~\ref{tab:halo_improv}, $b_2$ indeed is more tightly correlated with $b_1$ for that mass scale. The reason $b_\nabla$ is no longer as effective at predicting $b_\phi$ is two-fold: we mix many mass scales in the galaxy sample now so the concentration-$b_\phi$ relationship is weakened; and in addition, $b_{\nabla}$, which now absorbs a number of small-scale effects, is no longer tightly correlated with concentration.

\subsubsection{Best-fit parameters}

Ultimately, the goal of this work is to test the conjecture that pinpointing the Lagrangian bias parameters, which are measurable in a galaxy survey through e.g. the clustering statistics of a given galaxy sample, is beneficial to predicting $b_\phi$, and thus $\fnlloc$, with a smaller uncertainty. This seems to be manifestly true, as we are able to reduce the intrinsic scatter of $b_\phi$, from 0.64, 0.70 and 0.41 to 0.11, 0.12, and 0.09, respectively, for the three tracers (see Table~\ref{tab:gal_beta}), i.e. by about 80\% (see Table~\ref{tab:gal_improv}). For posterity, in Table~\ref{tab:gal_beta}, we provide the best-fit model parameter values that lead to this substantial improvement, i.e. the values of $\beta_0$ and $\beta_X$, the coefficients multiplying the Lagrangian parameters, $b_X$.
 
The $\beta$ coefficient values from this table along with the measured bias parameters $b_X$ can be used jointly to predict $b_\phi$ and serve as prior for a PNG analysis. It is reassuring that the best-fit $\beta$ coefficients for the two LRG samples are very similar, whereas the quasar sample yields slightly different values for $\beta$. We note that these values are specific to the analysis performed with the \textsc{AbacusSummit} suite and AbacusHOD prescription. In addition, when folding in our linear regression model into a PNG fit to real data, one would also need to marginalize over the error on these coefficients. Therefore, our recommendation is to perform the $b_\phi$ predictions on the fly with error bars when using \textsc{AbacusSummit} and otherwise, conduct a thorough recalibration of the fits. 

\section{Discussion and conclusions}
\label{sec:disc}

Inferring the value of the cosmological parameter $\fnlloc$ is crucial to understanding the origin of primordial fluctuations in the early Universe and ruling out or providing evidence for single-field inflationary models. Local-type PNG is usually investigated through the large-scale clustering properties of luminous tracers in cosmological surveys; however, to lowest order the quantity of interest, $\fnlloc$, is fully degenerate with the bias parameter, $b_\phi$, which characterizes the response of the particular tracer (e.g., galaxies, quasars) to changes in the fluctuations of the gravitational potential. While simplistic prescriptions such as the universality relation provide a link between linear bias and $b_\phi$, recent works have demonstrated that different tracers will exhibit considerable deviations from the universality relation, depending on the assembly history and other astrophysical properties of the galaxy sample. Budgeting this uncertainty into the $\fnlloc$ bounds is hefty, as $\sigma[\fnlloc] \propto \sigma[b_\phi]/b_\phi$, so any vestigial uncertainty on $b_\phi$ is also present in the constraints on $\fnlloc$. Additionally, adopting values for $b_\phi$ using the universality relation can also bias our inference on $\fnlloc$ for certain tracers. For this reason, finding alternative ways of calculating $b_\phi$ is of great importance to future surveys. This is precisely the objective of this paper.

In this work, we show that measuring the second-order Lagrangian bias parameters, $b_1$, $b_2$, $b_s$, and $b_\nabla$, in a galaxy survey is beneficial for predicting $b_\phi$ via a linear regression model, and thus, can be used for reducing the uncertainty on $\fnlloc$. We summarize our main findings as follows:
\begin{itemize}
\item Through Fig.~\ref{fig:halo_univ}, we demonstrate that we can obtain a good fit to the mean halo bias-$b_\phi$ relation by adopting a simple modification to the universality relation (see Eq.~\ref{eq:univ}). However, for samples preferentially featuring high- or low-concentration halos, we see a large variation in the $b_1$-$b_\phi$ relation. The same holds true for the rest of the second-order Lagrangian bias parameters, which also exhibit their own universality relations with $b_\phi$. These might be interesting to characterize in more detail in future works.
\item We examine the bias-$b_\phi$-concentration relation more closely for two representative halo mass bins in Fig.~\ref{fig:halo_ind}. Interestingly, we find that at fixed halo mass, the correlation between $b_\phi$ and the parameter $b_\nabla$ is strongest (with correlation strength over 90\%), followed by $b_1$. We attribute this to the finding that for a fixed mass halo sample, $b_\nabla$ turns out to be an excellent proxy for halo concentration, which captures the majority of the intrinsic $b_\phi$ scatter.
\item We follow up on this in Fig.~\ref{fig:halo_all} and Table~\ref{tab:halo_improv}, where we perform fits to each of the fixed halo mass samples using a simple linear regression model between the four Lagrangian bias parameters, $b_X$, and $b_\phi$. Visually, it is clear that there is a large amount of scatter in the values of $b_\phi$; however, we find that using even single-parameter models of the type $b_\phi = \beta_0 + b_X \beta_X$ can significantly reduce it. Which Lagrangian bias parameter to adopt in order to minimize the scatter depends on the mass of the halo sample: $b_\nabla$ yields a consistently large reduction, with $b_1$ and $b_2$ outperforming it only for small-mass halos. In all cases, the combined four-parameter model yields the best performance, allowing for more than 70\% smaller uncertainty on $b_\phi$. 
\item For maximum realism, we construct samples using the extended HOD model AbacusHOD (see Section~\ref{sec:hod}). Specifically, we study the three most suitable DESI tracer samples for PNG searches due to their large volume and high bias: LRGs at $z = 0.5$ and $z = 0.8$ and QSOs at $z = 1.4$. We construct a large suite of HOD samples for each of the three tracer samples by varying two mass-related, two concentration-related and two environment-related HOD parameters. As these samples mix a large range of halo masses, the effect on $b_\phi$ from varying the concentration is much smaller compared with the halos, whereas the response to the environment assembly bias parameters is notable.
\item Similarly to the halo samples, we study the reduction on the uncertainty of $b_\phi$ (and thus, indirectly, $\fnlloc$) when adopting a linear regression model with each of the Lagrangian bias parameters and all of them together. We show these as scatter plots with best-fit lines in Fig.~\ref{fig:gal_lrg1}, \ref{fig:gal_lrg2}, and \ref{fig:gal_qso}. We quantify the mitigation of the scatter in all of these cases in Table~\ref{tab:gal_improv}. Interestingly, we see that unlike the halo case, where some of the single-parameter models could significantly reduce the scatter, that is no longer the case for the galaxies, with the largest reduction being around 20\%. However, surprisingly and encouragingly, the full-parameter models yield much larger improvements, of over 80\% for all three tracers. We also provide the $\beta$ coefficients from our linear regression model for each tracer in Table~\ref{tab:gal_beta}, encouraging the use of our model in future searches of local-type PNG in DESI data.
\end{itemize}

While we show results for a specific HOD parametrization suitable for DESI galaxies, this work shows great promise for reducing the uncertainty on $b_\phi$ more broadly. Further studies are warranted to extend our results to more general galaxy populations, and comparing our results with similar analyses using hydrodynamical simulations and semi-analytical models will be very compelling \citep[as done in e.g.,][]{Shiferaw:2024ehr}.

We hope that future analyses of galaxy surveys can adopt Lagrangian-bias informed priors on the parameter $b_\phi$, such as those presented in this work, and be able to constrain the local-type PNG parameter $\fnlloc$ with less bias and smaller uncertainty.

\begin{acknowledgments}

We thank Stephen Chen, Nick Kokron, Elisabeth Krause, Uro{\v{s}} Seljak, James Sullivan, and Martin White for very helpful discussions and comments on the draft.
BH is supported by the Miller Institute for Basic Science.
SF is supported by Lawrence Berkeley National Laboratory and the Director, Office of Science, Office of High Energy Physics of the U.S. Department of Energy under Contract No.\ DE-AC02-05CH11231.

This research used resources of the National Energy Research Scientific Computing Center, which is supported by the Office of Science of the U.S. Department of Energy under Contract No.~DE-AC02-05CH11231.  Additional computations in this work were performed at facilities supported by the Scientific Computing Core at the Flatiron Institute, a division of the Simons Foundation.

\end{acknowledgments}

\section*{Data Availability}
The simulations used in this work are publicly available. Instructions for access and download are given at \url{https://abacussummit.readthedocs.io/en/latest/data-access.html}.

\bibliography{refs}{}
\bibliographystyle{prsty}



\end{document}